\documentclass[12pt]{article}
\usepackage{latexsym}
\newcommand{\beq}{\begin{equation}}
\newcommand{\eeq}{\end{equation}}
\newcommand{\bea}{\begin{eqnarray}}
\newcommand{\eea}{\end{eqnarray}}
\newcommand{\de}{\partial}
\newcommand{\f}{\phi}
\renewcommand{\a}{\alpha}
\renewcommand{\b}{\beta}
\renewcommand{\c}{\chi}

\newcommand{\w}{\wedge}
\renewcommand{\r}{\rho}

\newcommand{\g}{\gamma}
\newcommand{\e}{\epsilon}
\newcommand{\nn}{\nonumber}
\newcommand{\tr}{\mathrm{Tr}\,}
\renewcommand{\>}{\rangle}
\newcommand{\<}{\langle}
\newcommand{\om}{\omega}
\begin{document}
\baselineskip 18pt
\begin{titlepage}
\hfill hep-th/0311069
\vspace{20pt}

\begin{center}
{\large\bf{ HIDDEN sl(2,R) SYMMETRY IN 2D CFTs AND THE WAVE FUNCTION OF 3D 
QUANTUM GRAVITY}}
\end{center}

\vspace{6pt}

\begin{center}
{\large F. Nitti and M. Porrati} \vspace{20pt}

{\em Department of Physics\\ New York University\\ 4 Washington Place\\ 
New York NY 10003, USA}

\end{center}

\vspace{12pt}

\begin{center}
\textbf{Abstract}
\end{center}
\begin{quotation}\noindent
We show that all two-dimensional conformal field theories possess a hidden 
$sl(2,R)$ affine symmetry. More precisely, we add appropriate ghost fields to
an arbitrary CFT, and we use them to construct the currents
of $sl(2,R)$. We then define a BRST operator whose cohomology defines a 
physical subspace where the extended theory coincides with the original CFT.
We use the $sl(2,R)$ algebra to construct candidate wave functions for 3-d
quantum gravity coupled to matter, and we discuss their viability.
\end{quotation}
\vfill
 \hrule width 5.cm
\vskip 2.mm
{\small
\noindent e-mail: francesco.nitti@physics.nyu.edu, massimo.porrati@nyu.edu}
\end{titlepage}
\section{Introduction, Motivations (and Pipe Dreams)} 
Quantum gravity is still a mysterious theory, 
despite the enormous progress made in the last decades toward its 
understanding, mostly thanks to string theory. Even some
of the most basic questions are still unanswered. For instance, we do not 
have a clear understanding of what the fundamental, non-redundant 
degrees of freedom of quantum gravity are. One bold attempt to define them 
is the holographic principle~\cite{'t,s}. It states that the true degrees of 
freedom of quantum gravity in a region $V$ of a $d$-dimensional space can be 
thought of as describing a $d-1$-dimensional field theory, living at an 
appropriately defined boundary of the region $V$. The initial 
motivation for this idea is the famous Bekenstein bound~\cite{b} on the 
entropy of a black hole: $S=A/4G_N$ ($A=$ area of the black-hole horizon, 
$G_N=$ Newton's constant in $d$ dimensions). 

When the $d+1$-dimensional space-time is Anti de Sitter ($AdS_{d+1}$), 
this correspondence can be made
more concrete. In this case, the space-time boundary is a time-like 
surface; it is
conformal to the Einstein Static Universe, $S_{d-1}\times R$. 
In this case, the conjecture is that a consistent quantum gravity in an 
asymptotically $AdS_{d+1}$ space is equivalent (dual) to a 
{\em local, non-gravitational} field theory, living on $S_{d-1}\times R$.
The theory has a conformal fixed point in the ultraviolet (UV)~\cite{agmoo}. 
The equivalent theories are dual to each other, in the sense that when one of
them, say the field theory on $S_{d-1}\times R$ is strongly coupled, the
other one, quantum gravity on $AdS_{d+1}$, is weakly coupled (e.g. 
semi-classical).

Our description of the holographic duality has omitted many details. We should
selectively add them as needed in our discussion. The first one is
that the space gravity lives in, is not just $AdS_{d+1}$, but, generically,
a 10- or 11-dimensional manifold whose metric is a warped product of
$AdS_{d+1}$ times some compact space.
The first example of holographic duality, for instance, was between
Type IIB superstring theory on $AdS_5\times S_5$, and a $4$-d supersymmetric
Yang-Mills theory with 16 supercharges and gauge group $SU(N)$~\cite{m}.
The coupling constant of the $SU(N)$ super Yang-Mills theory (SYM) is
the 't Hooft coupling $g^2 N$. The dual meaning of this parameter is the 
curvature radius $l$ of $AdS_5$, in units of the string length $l_S$:
$l\sim (g^2N)^{1/4} l_S$~\cite{m}. Hence, the semi-classical regime for gravity
$l\gg l_S$, holds precisely when the SYM theory is strongly coupled.
In this example, the $AdS_5\times S_5$ background is a near-horizon geometry
of $N$ parallel D3 branes.

Other examples of $AdS$  holography have been worked out in the 
literature.
A particularly interesting one gives origin to an $AdS_3/CFT_2$ 
duality~\cite{sm,db}.
In this case, one starts with a configuration of $Q_1$ D1 branes and $Q_5$
D5 branes of Type IIB superstring on $R^{(5,1)}\times M^4$, with $M^4$ either
$T^4$ or $K_3$. The near-horizon geometry of this configuration is
$AdS_3 \times S_3 \times M^4$. The holographic dual is the 
infrared limit of a 2-dimensional conformal field theory, which  
has central charge $c=Q_1Q_5$ [up to $O(1)$ corrections] and is believed 
to be a deformation of the orbifold sigma models living on the symmetric 
product of $c$ copies of $M^4$. The curvature radius of $AdS_3$ is 
$l\sim (Q_1Q_5)^{1/4}G_N$, where $G_N$ is the 3-dimensional Newton 
constant.

More generally, one can conjecture that any {\em consistent} quantum gravity
on $AdS_3$ is described by some conformal field theory. 
One important clue to this conjecture is that the algebra of
asymptotic isometries in $AdS_3$ gravity is the Virasoro algebra with
central charge $c= 3l/2G_N$~\cite{B&H}. 

A special case is pure $AdS_3$ gravity. This theory does not propagate any
local degrees of freedom, so, by construction, it has only a boundary 
dynamics. We shall explain in details in Section~\ref{ads3 grav} how to
rewrite pure (2+1)-dimensional gravity as a  Chern-Simons (CS) theory
with gauge group $SL(2,R)\times SL(2,R)$. Here we only remark that any CS
theory is ``hyper-holographic.'' By this we mean the following.

\subsection{Hyperholography}

Euclidean $AdS_{d+1}$ space is topologically a $d+1$-dimensional ball, 
$B_{d+1}$. The space can 
be foliated by spheres $S_d$, with coordinates $x^\mu$, 
together with a radial coordinate $r$. Near the 
boundary $r=0$, the line element is
\beq\label{m1}
ds^2\approx {l^2\over r^2} [dr^2 + g_{\mu\nu}(r,x)dx^\mu dx^\nu], \qquad
g_{\mu\nu}(r,x)= g^0_{\mu\nu}(x) + O(r^2).
\eeq 
In radial
quantization, $r$ plays the role of time: to describe quantum gravity, 
one gives the wave function ``of the universe'' at fixed $r$, and dynamics is
radial evolution.
The AdS/CFT duality becomes the statement that the wave function of the 
universe at $r=\epsilon \ll l$ 
is the partition function of a CFT regularized with
cutoff $\epsilon$~\cite{m2}:
\beq\label{m2}
\Psi(\epsilon)=Z_{CFT}^\epsilon.
\eeq
The cutoff $\epsilon$ must be introduced because the partition function of the
CFT, $Z^\epsilon_{CFT}$, may (does) contain divergent contact terms.
More concretely, if we make explicit the functional dependence of 
$\Psi_\epsilon$ on the metric we have
\beq\label{m3}
\Psi[\epsilon, g_{\mu\nu}]=\exp \{-\int d^dx [A\epsilon^{-d} + 
B \epsilon^{2-d}R(g) +... ] - \Gamma^F_\epsilon [g]\}. 
\eeq
The ellipsis denote divergent, {\em local} functions of the metric, $A,B$ are
constants, and $\Gamma^F_\epsilon (g)$ is a (non-local) functional, finite
in the limit $\epsilon\rightarrow 0$.

CS theories obey an identity stronger than Eq.~(\ref{m3}). In that case, 
given any
3-dimensional space $M$ with boundary $\Sigma=\partial M$, one 
has~\cite{w1}
\beq\label{m4}
\Psi_{CS,k}[\Sigma]=Z^\Sigma_{WZW,k}.
\eeq
In words: on {\em any} surface $\Sigma$, 
the wave function of the CS theory with gauge group $G$ and coupling constant
$k$ 
{\em is} the partition function of the level-$k$ chiral 
WZW model of the group $G$.
This definition can be easily modified so that it makes  sense also when 
the surface $\Sigma$ has a boundary. In this case, it is no longer true that 
$\Sigma=\de M$, rather: $\Sigma \subset \Sigma'=\de M$. 
Indeed, Eq.~(\ref{m4}) makes sense for a larger
class of 2-dimensional CFTs, as we will argue now.

\subsection{A Wave Function of the Universe?}

The identity Eq.~(\ref{m4}) holds because both sides of the equation
obey the same defining functional equation: the LHS obeys a quantum 
Gauss law, while 
the RHS obeys the Ward identities of the affine-Lie algebra based on the 
group $G$ \cite{seiberg}.
Now, the Gauss law is the defining equation for the wave function, so we may
ask whether one can get a reasonable Gauss law~\footnote{For gravity, the 
quantum Gauss law is of course the Wheeler-De Witt equation.}
from other 2-d CFT, besides the chiral WZW model.
Consider in particular a generic 
2-d CFT possessing holomorphic dimension-1 currents 
$J_a(z)$, which generate the affine-Lie algebra $\hat{g}$ based on the 
group $G$. They obey the well-known OPE
\beq\label{m5}
J_a(z)J_b(w)= {k \delta_{ab} \over (z-w)^2} + {i f_{ab}^c \over (z-w)} J_c(z) +
\mbox{regular terms}.
\eeq
The $f_{ab}^c$ are the structure constants of the group, and $k$ is the level 
of the affine algebra.
Consider now the (genus-0) partition function of this model, in the presence of
a background gauge field $A_{\bar{z}}^a$. This is also the generating functional for
the Green's functions of the currents $J_a$:
\beq\label{m6}
Z[A_{\bar{z}}]\equiv \langle 0 | \exp \int d^2 z J_a A_{\bar{z}}^a |0\rangle .
\eeq
On the background $A_{\bar{z}}^a$, the current is covariantly conserved up 
to an anomaly
\beq\label{m7}
(D_{\bar{z}} J)^a= {ik \over 2\pi} \partial_z A^a_{\bar{z}}, \qquad 
D^a_{\bar{z}\,b}=\delta^a_b \partial_{\bar{z}} +f_{bc}^a A^c_{\bar{z}} .
\eeq
Substituting this equation into the definition Eq.~(\ref{m6}), we get
\beq\label{m8}
\left[D_{\bar{z}} {\delta \over \delta A_{\bar{z}}^a(z)}- 
{ik \over 2\pi} \partial_z A^a_{\bar{z}}(z) \right] Z[A_{\bar{z}}] = 0.
\eeq
If we interpret $Z[A_{\bar{z}}]$ as a wave function, and we define the 
canonically  conjugate variable to $A_{\bar{z}}^a$ by 
\beq\label{m9}
A_{z\,a}(z,{\bar{z}}) \equiv -{2\pi i\over k} 
{\delta \over \delta A_{\bar{z}}^a(z,\bar{z})},
\eeq
then, Eq.~(\ref{m8}) becomes the Gauss law $F_{z\bar{z}}^a Z[A_{\bar{z}}]=0$.

An obvious generalization is to insert local operators in Eq.~(\ref{m6})
\beq\label{m10}
Z_{\cal O}[A_{\bar{z}}]= \langle 0 | R \prod_i {\cal O}^i(z_i)
\exp \int d^2 z J_a A_{\bar{z}}^a |0 \rangle, \qquad \mbox{$R$= radial ordering}.
\eeq
The operators ${\cal O}^i$ may have a singular OPE with $J^a$. When they are 
primaries of the affine-Lie algebra, the OPE is particularly simple
\beq\label{m11}
J^a(z){\cal O}^i(w)={1\over z-w} T^{a\, i}_j {\cal O}^j.
\eeq
The matrices $T^{a\, i}_j$ define a representation of the group $G$.

The anomalous Ward identity now gives the Gauss law with external point-like
charges
\beq\label{m12}
F_{z{\bar{z}}}^a Z_{\cal O}[A_{\bar{z}}] = \sum_i k\delta^2(z-z_i)  
\langle 0 | R \prod_{l\neq i} {\cal O}^l(z_l) T^{a\, i}_j {\cal O}^j(z_i)
\exp \int d^2 z J_a A_{\bar{z}}^a |0 \rangle.
\eeq

More generally, one can define a wave function by considering the CFT on a 
disk $D$, with boundary state $|B\rangle$ at $\partial D$.
\beq\label{m13}
Z_{{\cal O},B}[A_{\bar{z}}]=
\langle 0 | R \prod_i {\cal O}^i(z_i)\exp \int_D d^2 z J_a A_{\bar{z}}^a 
|B\rangle.
\eeq
For any point $z\in D-\partial D$, $Z_{{\cal O},B}[A_{\bar{z}}]$ 
obeys the Gauss law
Eq.~(\ref{m12}).

Equation~(\ref{m12}) is suggestive. If we take it literally, it says that
the partition function of {\em any} CFT possessing an 
affine-Lie algebra can 
be reinterpreted as the physical wave function of a CS theory coupled to 
matter. Notice that the CFT need not be a WZW model.
Furthermore, as we will review in Section 2, pure gravity in 3-d is a CS
theory with gauge group $SL(2,R)\times SL(2,R)$. So, Eq.~(\ref{m12})
or (\ref{m13}) may be used to define the Wheeler-De Witt wave function.

\subsection{Plan of the Paper}

Our last statement was quite imprecise. There are many subtle points to
understand before we can, even tentatively, identify Eq.~(\ref{m13})
with the wave function 
of 3-d AdS gravity. In Section 2 we review the construction of CS
3-d gravity, and its reduction to a boundary WZW model with constraints. 
In the course of the review, we address 
the first subtlety. The problem is that, 
on manifolds with boundaries, pure 3-d gravity needs an extra constraint on the
space of states. Section 2 shows how to modify 
accordingly the recipe for the physical wave function. 

The second problem is whether an affine-Lie symmetry $sl(2,R)$ exists,
inside a generic CFT. Better, whether, given an arbitrary CFT, one
can extend it in such a way that:
a) The extended theory possesses an affine $sl(2,R)$.
b) One can define a physical subspace within the extended theory, where it 
reduces to the original CFT.
c) The stress-energy tensor of the extended theory coincides with that of the
original theory on physical states.
d) The physical subspace is defined by a BRST cohomology.
e) Any physical operator of the original CFT can be ``dressed'' appropriately
so that it transforms in a representation of the affine $sl(2,R)$.
In Sections 3 to 7 we show that all these conditions can be satisfied.

Section 3 answers the first question. There, a minimal set of auxiliary 
ghost fields is defined, that allow us to construct the currents of
$sl(2,R)$. Our construction is a modification of the well-known 
free-field realization of the $sl(2,R)$ current 
algebra.
Section 4 uses a known technique for modifying the stress-energy 
tensor, and consistently implement the extra constraint
needed to reduce CS states to physical, 2-d gravity states. 

Section 5 
introduces the BRST operator that defines the physical space, and shows that
the stress-energy tensor of the extended theory equals that of the original 
theory, up to a BRST-exact term, thereby answering question c). Points b) and
d) are answered in Section 6, where the BRST cohomology is proved to coincide 
with the Hilbert space of the original CFT. Section 7 addresses the last point:
it gives a constructive recipe to build irreducible representations of the
affine $sl(2,R)$ starting from the Virasoro primaries of the original CFT.

Sections 3 to 7 are the converse of ref.~\cite{B&O}. There, it was shown that
an irreducible representation of affine $sl(2,R)$ can be constrained so as to 
give a single, irreducible representation of the Virasoro algebra (see 
ref.~\cite{superB&O} for the supersymmetric extension of this result).
Here, we show how to embed a CFT, i.e. a collection of representations of the
Virasoro algebra, into a collection of representations of $sl(2,R)$. 

The most serious problem toward using Eq.~(\ref{m13}) to
define a physical wave function for 3-d gravity arises precisely in the 
presence of matter. In that case, the Gauss law becomes $F_{z\bar{z}}^a\Psi=
\rho^a\Psi$, where $\rho^a$ is the charge density of $SL(2,R)$. The problem is
that this charge density must also be the stress-energy tensor of matter.
Whether $\rho^a$ gives an acceptable stress-energy tensor is an 
open problem. This point is discussed more extensively in Section 8, 
which also contains our conclusions. 

Technical material on $AdS$ boundary 
conditions is confined to Appendix A. Appendix B gives a derivation of the
well-known K\"unneth formula in BRST cohomology, and is included for 
completeness. 
 
\section{$sl(2,R)$ Affine Lie Symmetry in $(2+1)$-Dimensional 
\newline Gravity}\label{ads3 grav}

In this Section  we review the $sl(2,R)$ affine algebra structure of gravity
in (2+1) dimensions with a negative cosmological constant $\Lambda=-1/l^2$,  and
how $AdS_3$ boundary conditions determine the constraints on the affine
currents, following \cite{chvd,banados}.

\subsection{3-d Gravity as CS}

Einstein gravity in (2+1) dimensions with a negative cosmological constant can
be reformulated in terms of two copies of an $SL(2,R)$ CS  
gauge
theory~\cite{w,Achucarro:vz,tv}. With the definitions
\beq\label{gf}
A^a = {e^a \over l}+ \omega^a  , \qquad \qquad \tilde{A}^a = -{e^a \over l} +
\omega^a,
\eeq
the Einstein-Hilbert action becomes
\beq
S_E \equiv{1 \over 8\pi G_N} \int \left(e^a \wedge R_a - {1\over 6\,l^2}
\e_{abc}\,e^a\wedge e^b\wedge e^c\right)  = S_{CS,k}[A] -  S_{CS,k}[\tilde{A}].
\label{cs}
\eeq
Here $S_{CS,k}[A]$ is the Chern-Simons action with coupling constant $k$ for the $SL(2,R)$  gauge connection
$A=A^a t^a$, $t^a$ are
the generators of the $sl(2,R)$ Lie algebra in the fundamental representation,
and $k=-l/4G_N$. The topological character of the CS action implies that there
are no local degrees of freedom in this theory, and the dynamics is given
entirely in terms of holonomies of the (flat) gauge connections $A$ and
$\tilde{A}$.
Things change however if the 3-d manifold on which the theory is defined has a
time-like boundary: in this case the
CS gauge theory has  an infinite number of degrees of freedom, consisting 
 of the values of the  gauge fields at the boundary, and  the model
describing the boundary dynamics possesses  an affine-Lie algebra structure,
based on the group $SL(2,R)$. \\
On the other hand, if we consider the metric description of $AdS_3$ gravity, we
must demand that the metric approaches
the  $AdS_3$ metric near the boundary. Using coordinates 
$(r,x^+=t/l+\phi,x^-=-t/l+\phi)$, this metric reads
\beq \label{Ads bc}
d s^2 \approx l^2 \left(d r^2  + e^{2 r} d x^+ d x^-\right), \qquad \qquad r
\rightarrow \infty.
\eeq
This condition is obeyed only by a restricted class of CS
connections; therefore, the theory describing the dynamics of asymptotically
$AdS_3$ spaces, will not possess the full affine-Lie symmetry of the boundary
CS, but only a subgroup
preserving the boundary conditions Eq.~(\ref{Ads bc}). This subgroup, made of
two copies of the Virasoro algebra, was
found long ago~\cite{B&H} to be the asymptotic symmetry group of $AdS_3$. One
can enforce this restriction on the dynamical degrees of freedom  through a
mechanism called Hamiltonian reduction. It can be realized by gauging part of 
the full algebra, so that only a subalgebra connects physically inequivalent
states. This mechanism is the main
subject of the next Sections. In this Section, we review the constraints that
$AdS_3$ boundary conditions impose on the
affine currents.

Consider the bulk CS action\footnote{To avoid repetitions,
we consider here only one of the two CS theories in Eq.~(\ref{cs}).}
\beq\label{CS}
S_{CS,k}^0[A] \equiv {k \over 4\pi}\mathrm{Tr}\,\int_M  \left (A\wedge d A + {2
\over 3} A \w A \w A \right) =  {k \over 4\pi}\mathrm{Tr}\, \int_M \e^{\mu\nu\r}\left
(A_\mu\de_\nu A_\r + {2 \over 3} A_\mu  A_\nu  A_\r \right),
\eeq
where $M$ is a 3-manifold with coordinates $(r,t,\phi)$, whose boundary is
parametrized by $(t,\phi)$, or $(x^+, x^-)$.   To have
a well defined variation, this action needs boundary conditions that fix one of
the components of the gauge field $A$. Different choices of boundary conditions
can be implemented by adding appropriate boundary terms to Eq.~(\ref{CS}), and
demanding that the action is stationary with respect to
{\em all} smooth variations of the fields, even those that do not vanish at 
the boundary. 
By adding the boundary term $\mathrm{Tr}\,\int_{\de M} A_t A_\f$
to Eq.~(\ref{CS}), one brings the action in a ``canonical'' form, 
where $A_t$ appears explicitly as a Lagrange multiplier:
\beq\label{CScan}
S^1_{CS,k}[A]\equiv S_{CS,k}^0[A]\,+\,{k \over 4\pi}\mathrm{Tr}\,\int_{\de M} A_t A_\f =
 {k \over 4\pi}\tr \int_M \left(A_\phi \de_t A_r - A_r \de_t A_\f + 2 A_t F_{r
\f}\right).
\eeq
The variation of this action w.r.t. $A_t$ yields the constraint $F_{r \f}=0$,
while the variation w.r.t. the other components of $A$ gives terms
proportional to the equations of motions $F_{t r} = F_{t \f} = 0 $, plus
the boundary term $(k/4\pi)\tr \int_{\de M} A_t \delta A_\f$. To make it
vanish we must require the boundary condition $A_t=0$. 
However, as we review in Appendix A, the boundary
conditions for asymptotically $AdS_3$ space-times are $A_- =
0$, or $A_t = A_\f$. To enforce them, we add to Eq.~(\ref{CScan}) an additional
boundary term, and use the following definition of the CS action:
\beq\label{CSbc}
S_{CS,k}[A] \equiv {k\over 4\pi}\tr \int_M \left(A_\phi \de_t A_r - A_\r \de_t
A_\f + 2 A_t F_{r \f}\right)\, +\, {k \over 4\pi}\tr \int_{\de M} A_\f^2.
\eeq 
The constraint $F_{r \f}=0$ is solved by the requirement that the
space part of the connection is flat. On a disk without punctures this implies
\beq \label{flat}
A_r = U \de_r U^{-1}, \qquad \qquad    A_\phi = U \de_\phi U^{-1},
\eeq
with $U(t,r,\phi)$ an arbitrary element of $SL(2,R)$. Substituting 
Eq.~(\ref{flat}) back into Eq.~(\ref{CSbc}), and integrating by parts,  
we get an
induced action for the group element $U$, which reads 
\bea\label{CWZW}
S^+[U] &=& 
{k \over 4\pi} \tr \int_{\de M}  \left[\left(U \de_t U^{-1}\right)\left(  U \de_\phi
U^{-1}\right) -  \left(U \de_\phi U^{-1}\right)^2\right] d t d \phi \nn\\
 && \,+\, {k \over 12\pi} \tr \int_{M} \e^{\mu\nu\rho}  \left[\left(U \de_\mu
U^{-1}\right)\left(U \de_\nu U^{-1}\right)\left(U \de_\rho U^{-1}
\right)\right] d^3
x.
\eea
This is the chiral WZW action~\cite{sonnenschein,stone} for
$U$. It is 2-dimensional, since it depends only on the boundary value of
$U$: $U(t,\phi)$. It has an affine $SL(2,R)$ symmetry of
the form $U \to h(x^+)U$,  generated by the right-moving current $J(x^+) =k\,
U\de_\f U^{-1} =k\, A_\f =k\, A_+/2$. There is only a
right-moving affine Lie symmetry because the boundary condition
$A_-=0$ is preserved only by gauge transformations on $A$ that are 
independent of $x^-$ at the boundary.

The same derivation can be carried out for the other CS theory, 
with connection $\tilde{A}$: for this we impose the boundary condition 
$\tilde{A}_+=0$, leading to
a chiral WZW model on the boundary, with a left-moving affine-Lie symmetry
generated by the current $\tilde{J}(x^-) = k\,\tilde{A}_-/2$. This shows that
the boundary degrees of freedom of gravity in (2+1)-dimensions realize two
independent chiral affine Lie algebras.

Up to now, we have imposed only a minimal set of boundary conditions on
the CS gauge fields, 
just enough to make the variation of the action
well defined. If we want the CS theory to describe gravity in an asymptotically
$AdS_3$ space-time, we need to put further restrictions on the boundary values
of $A$ and $\tilde{A}$. To see this, recall that in an asymptotically $AdS_3$
space-time, the metric near the boundary must reduce to the form given in
Eq. (\ref{Ads bc}). More precisely, the metric must have the asymptotic
behavior \cite{B&H}:
\bea
d s^2 &=& l^2 \left[\left(1 + O(e^{-2 r})\right) d r^2  \,+ \,\left(e^{2
r}+O(1)\right) d x^+ d x^-\right. \nn\\
&+& \left.O(1)(d x^+)^2 \,+\, O(1)(d x^-)^2 + O(e^{-2 r})(d rd x^+ +
d rd x^-)\right].
\eea
As shown in Appendix A, this asymptotic behavior not only  requires the boundary
condition\footnote{The upper index refers the $sl(2,R)$ Lie algebra, the lower
one to space-time coordinates}
\beq\label{mm0}
A_-=0, \qquad \qquad \tilde{A}_+ = 0,
\eeq
but it also constrains the WZW affine currents to satisfy
\beq \label{j-constraint}
J^- = k, \qquad  J^3 = 0; \qquad\qquad  \tilde{J}^+ = -k,\qquad  \tilde{J}^3=0,
\eeq
with arbitrary $J^+$ and $\tilde{J}^-$.

Clearly this restriction on the boundary values of the fields  breaks the 
affine-Lie symmetry; as we will see shortly, 
it leaves only a conformal symmetry of the
boundary, generated by two independent  copies of the Virasoro algebra.
This can be  shown by 
imposing $J^-=k$ and $J^3=0$ as classical constraints in the boundary theory,
and observing that the Dirac brackets of the only remaining field, 
$J^+(x^+,x^-)=k L(x^+,x^-)$, are precisely those of a Virasoro
algebra with central charge $c=-6k$~\cite{banados}. Another approach
\cite{chvd,HMS} consists in further reducing the action (\ref{CWZW}) by imposing
the constraints {\em after} 
having combined the two chiral WZW models in a single
non-chiral one, with group element $g^{-1}(x^+) \tilde{g}(x^-)$. As it 
was  shown in ref.~\cite{Forgacs:ac} the result of this reduction  is
the Liouville theory, which carries an action of the left and right Virasoro
algebras.
We will not  follow these  approaches. Instead, we will impose 
$J^-=k$ {\em after} 
quantization, as a constraint on physical states. 
In this approach, i.e. if we first quantize and then
impose the constraint, the condition  $J^-=k$ is enough to get rid of the
unphysical degrees of freedom, whereas the condition $J^3=0$ is a
gauge-fixing. 

Until now we have discussed pure gravity, but one expects this discussion
to be valid also for gravity coupled to arbitrary  sources
localized in the bulk.
In fact, the Virasoro algebra is an asymptotic isometry of $AdS_3$ 
even when matter is added to pure 3-d gravity~\cite{hmtz}.
This is one of the basis for the strong version of the 
$AdS/CFT$ correspondence  in 2+1 dimensions, which conjectures that 
any {\it consistent} theory of quantum gravity in an
asymptotically $AdS_3$ space-time is dual to some 
2-d CFT living on the boundary. 
One may wonder if this result is valid not only for the conformal structure,
but also for the affine-Lie structure, i.e. if, for a generic theory of $AdS_3$
gravity, one can construct affine currents acting on the boundary, which, upon
restriction to $AdS_3$ boundary conditions, reduce to the Virasoro algebra. It
would be very hard to check this fact along the lines  followed in this
Section. Indeed, the CS description is simple 
only in the case when matter is made of point-like external 
sources in the bulk.
In this case, sources are represented by punctures in the disk, together 
with their associated gauge-field holonomies. For a general matter 
QFT coupled to gravity, instead, it is generically impossible to reformulate 
the model as an ordinary  gauge theory: the kinetic terms of the matter 
involve the inverse of the dreibein, so that
the coupling to the CS gauge field is very complicated, and certainly
non-minimal. 

The presence of the boundary affine-Lie symmetry, instead, follows
from a \emph{purely 2-d} result. 
This is the main point of this paper: we will
show in  Sections 3 to 7  that all 2-d CFTs possess 
a ``hidden'' $sl(2,R)$ affine algebra, namely,
that it is always possible to embed the Virasoro algebra in a larger
 $sl(2,R)$ affine structure, by adding appropriate auxiliary fields. We will
also define a physical subspace --and physical observables-- such that the
restriction of the  new theory to that physical subspace gives back
the original theory and the original Virasoro algebra. Before doing that, we
make some remarks about how the constraints on the current are imposed on the 
wave function. 

\subsection{The Extra Constraint}

Before describing the construction of the hidden $sl(2,R)$ algebra, we must 
show how the extra constraint 
changes the definition of the wave function Eq.~(\ref{m13}).  Clearly, 
nothing changes for closed 2-d manifolds. When the 2-d manifold has a 
non-empty boundary, 
instead, the extra constraint imposes a restriction on the 
boundary state. In the simple case considered in Eq.~(\ref{m13}), i.e. 
a disk with punctures, the constraint on the state $|B\rangle$ is
\beq\label{mm1}
(J^-_n -k\delta_{n,0})|B\rangle=(\tilde{J}^+_n  + k\delta_{n,0})|B\rangle=0, \qquad \forall n,
\eeq
where $J^-_n$ are the modes of the current $J^-(z)$.
In later Sections we will recast this constraint in a BRST form 
$Q_{B}|B\rangle=0$. In either form, Eq~(\ref{mm1})  implies a set of Ward 
identities on the wave function. Consider for instance the constraint
following from $J^-=k$:
\beq\label{mm2} 
\langle 0 | R \oint_{C=\partial D} {dw\over 2\pi i} 
w^{n} [J^-(w)-k] \prod_i {\cal O}^i(z_i)
\exp \int_D d^2 z J_a A_{\bar{z}}^a |B\rangle=0.
\eeq
Call $D_i$ an infinitesimally small disk centered around the $i$-th puncture. 
Define
$C_i=\partial D_i$ and $\delta_n{\cal O}^i(z_i)=\oint_{C_i} {dw\over 2\pi i} 
w^n[J^-(w)-k]{\cal O}^i(z_i)$.  
By deforming the contour of integration $C$ past all the operator insertions,
and using the OPE of the affine-Lie algebra, we find
\beq\label{mm3}
\sum_i\langle 0 | R \prod_{l\neq i} {\cal O}^l(z_l) \delta_n{\cal O}^i(z_i)
\exp \int_D d^2 z J_a A_{\bar{z}}^a |B\rangle +
\left[\int_{D-\sum_i D_i} d^2 z z^n F_{z\bar{z}}^+  + \oint_{C-\sum_i C_i} dz z^n 
A^+_{\bar{z}}\right]Z_{B\,{\cal O}}=0.
\eeq
The first term in brackets vanishes because of Gauss law, while the second
vanishes on any smooth gauge field configuration obeying the
asymptotic condition Eq.~(\ref{mm0})\footnote{Recall that $A_{\bar{z}}$ 
corresponds to  $A_-$ in the euclidean theory.}, so we arrive at the constraint
\beq\label{mm4}
\sum_i\langle 0 | R \prod_{l\neq i} {\cal O}^l(z_l) \delta_n{\cal O}^i(z_i)
\exp \int_D d^2 z J_a A_{\bar{z}}^a |B\rangle =0.
\eeq
In the BRST formalism, the constraint translates into
\beq\label{mm5}
\sum_i \langle 0 |R \prod_{l\neq i} {\cal O}^l(z_l) [Q_B,{\cal O}^i(z_i)]
\exp \int_D d^2 z J_a A_{\bar{z}}^a |B\rangle =0.
\eeq
In particular, it is satisfied if the operators ${\cal O}^i(z_i)$ are BRST-invariant. 

\section{The $sl(2,R)$ Currents}

In this Section we show that, given a generic 2-dimensional  CFT, it is always
possible to
add a universal  set of auxiliary free fields such that the resulting theory carries an
affine $sl(2,R)$
algebra structure. Our construction is inspired by free field 
realizations of affine-Lie algebras first introduced 
in~\cite{Wakimoto,zamol}, further generalized and analyzed 
in~\cite{B&O,Feigin,gerasimov,Bernard:1989iy,bouwknegt,ohta,Furlan} 
(see \cite{bouwknegtrev} for a review and further references).
From now on we switch to Euclidean notation
and we use  holomorphic-antiholomorphic (rather than left-right moving)
coordinates. 

We  start with a CFT  (henceforth referred to as the ``matter,'' or
``physical'' CFT) with stress tensor $T_m$ and central charge $c_m$, satisfying
the OPE
\beq
T_m (z)T_m(w) \sim \frac{c_m/2}{(z-w)^4} + \frac{2T_m(w)}{(z-w)^2} + \frac{\de
T_m(w)}{z-w},
\eeq
but otherwise generic. In particular, we 
do not make any assumption regarding the spectrum of the operators, or the 
nature of the interactions.
We can regard this CFT as the boundary theory, describing some generic matter
coupled to gravity in
$AdS_3$. Then, the stress tensor $T_m$ represents the generator of the Virasoro
algebra of asymptotic symmetries of the  $AdS_3$ theory.
As a first step in the construction, we 
add the following set of auxiliary free fields: two scalars fields  $\rho(z)$
and $\chi(z)$, and a pair of bosonic ghost fields $(\b,\g)$ of weight $(0,1)$.
The scalar field $\rho$ is a ``ghost'': its kinetic term has opposite sign 
compared to that of a physical scalar. These fields have  OPEs
\beq\label{OPEs}
\b(z)\g(w) \sim \frac{1}{z-w},  \qquad \c(z)\c(w)  \sim \ln \frac{1}{z-w},
\qquad \rho(z)\rho(w) \sim \ln (z-w).
\eeq
Furthermore, we assume that 
the field $\c$ has a background charge $\a_\c$,  such that
the central charge
in the $\c$-sector, $c_\c\equiv 1-12\a_\c^2 $ , is equal and opposite to that
of the physical sector. So, we take for $\c$ a (non-unitary)
stress-energy tensor of the form
\bea
&&T_\c=-\frac{1}{2}(\de \c)^2 +i \a_\c \de^2 \c, \\
&& 1-12\a_\c^2=-c_m \label{chi}.
\eea
Next, we define the $sl(2,R)$ currents
\bea \label{currents}
J^+(z) & = & -\b(z)[\g(z)]^2  - {\a}_{+} \g(z) \de \rho (z) + k \de \g(z) +
(k+2)[\beta(z)]^{-1}\,\Big[T_m(z) + T_{\c}(z)\Big], \\
J^3(z) & = & \b(z)\g(z) + \frac{1}{2}\a_+ \de \r(z),\\
J^-(z) & = & \b(z),
\eea
where $\a_+=\sqrt{2k+4}$. These currents are similar to the ones appearing in  
the standard  free-field realization of the $sl(2,R)$ 
affine algebra as it appears in~\cite{B&O}, the 
 difference being the presence of the last term in 
$J^+(z)$\footnote{The presence of an inverse
power of the ghost field $\b(z)$ in the definition of $J^+$ may seem unusual,
and one may worry that it is ill-defined. In general, one can define an 
arbitrary power of a field through its OPE with other fields,
as is done in the context of ``fractional calculus'' (see e.g.
\cite{Rasmussen} and references therein). Alternatively, one can ``bosonize''
the $(\b,\g)$ pair by trading it for two scalar fields $\phi(z)$, $\psi(z)$,
with OPEs $\phi(z)\phi(w)\sim -\ln(z-w)$, $\psi(z)\psi(w)\sim \ln(z-w)$,  and
the identifications
\begin{displaymath}
\b=\exp (\phi - \psi),   \qquad \g = \de \psi \exp (\psi - \phi). 
\end{displaymath}
Then one has $\b^{-1} \equiv \exp (\psi - \phi)$.}.  
As a consequence of  the OPEs in Eq.~(\ref{OPEs}), 
and of the stress-energy tensor OPE, 
these operators satisfy the $sl(2,R)$ affine-Lie
algebra at level $k$:
\bea
J^+(z) J^-(w) &\sim& \frac{2 J^3(w)}{z-w} + \frac{k}{(z-w)^2}, \nn \\
J^3(z) J^+(w) &\sim& \frac{J^+ (w)}{z-w}, \qquad J^3(z) J^-(w) \sim -\frac{
J^-(w)}{z-w}, \nn \\
 J^3(z)J^3(w) &\sim& \frac{k/2}{(z-w)^2}, \qquad J^+(z)J^+(w) \sim 0,  \qquad
J^-(z)J^-(w) \sim 0.
\eea
The coefficient of the last term in $J^+$ is fixed by demanding that
$J^+(z)J^+(w) \sim 0$. This also imposes the requirement that 
$T_m +T_\c$ has vanishing central charge. 
Up to  this point, the level $k$ is arbitrary, and  not related
to the value of the central charge of the physical theory. However, we shall 
see that, if we require  this extended theory to be physically equivalent to the CFT we
started with, the value of $k$ will be uniquely determined in terms of $c_m$. 

The standard Sugawara stress tensor associated to the $sl(2,R)$ algebra is
\bea
T^{sug} &=& \frac{1}{2(k+2)}\left(:J^+J^-:+:J^-J^+:+\,2:J^3J^3:\right) \nn \\
& = &  \b \de \g + \frac{1}{2} (\de \r)^2 + \frac{1}{ \a_+} \de^2 \r + T_\c +
T_m, 
\eea
where $:\, :$ denotes the normal ordering.  This stress-energy tensor 
has central charge $c_{{\textrm \tiny{SL_2}}}= 3 k /(k+2)$. Under $T^{sug}$,  
the currents in Eq.~(\ref{currents}) are primary operators of weight one.
Notice that the field $\r$ has a background charge $1/\a_+$, so, its
contribution to the central charge is $c_\r = 1-12 \a_+^{-2} = (k-4)/(k+2)$.
In the next Section, we shall see that, to be able to impose 
the constraints discussed in Section 2, we will
have to change the conformal weight of $J^-$, by adding an improvement term
to $T^{sug}$.

\section{Constraints}

As we have seen in Section \ref{ads3 grav}, imposing $AdS_3$ boundary
conditions is equivalent to imposing appropriate constraints on the $sl(2,R)$
currents, since the asymptotic isometries of $AdS_3$ generate only the Virasoro
algebra, rather than the full affine-Lie algebra. This is true for
pure gravity as well as 
for gravity coupled to matter, as long as the matter fields
have a boundary behavior that does not spoil the asymptotic form of the
metric.  To reduce the full $sl(2,R)$ symmetry to  the Virasoro algebra, 
we need to impose the constraint
Eq.~(\ref{j-constraint}), $J^-(z) = k$. Although the reduced theory has only a
Virasoro symmetry, this constraint does not remove all auxiliary fields from
the spectrum, since it does not act on the fields $\r$ and $\c$. 
To eliminate these extra fields, and to reduce the current algebra to the 
\emph{physical} Virasoro algebra associated to $T_m$, 
we need to impose some additional constraints. It
turns out that the extra condition  $\de\r =  \de\c$ is sufficient to 
our purpose.

We are going to impose these constraints in a consistent way, 
 using the BRST formalism, 
in the next  Sections. Here we want to give a  heuristic idea of
how  these constraints, together with a condition on $k$ , give the stress
tensor $T_m$, and the correct central charge $c_m$, starting from $T^{sug}$ 
and its central charge
$c_{\textrm \tiny{SL_2}}$.

First of all, notice that  the constraint $J^-(z) = k$ is meaningless if
$J^-(z)$ is a field of dimension one. This is clear from dimensional analysis.
Equivalently, this constraint does not commute with the Sugawara Hamiltonian,
$L_0^{sug}= (2\pi i)^{-1} \oint d z \,z T^{sug}(z)$:
\beq
\left[L_0^{sug}, J^-(w)\right] = J^-(w) + w \de J^-(w).
\eeq
From this equation, we can see that that difficulty is  overcome if we
modify the stress tensor in such a way that $J^-(z)$ becomes a field of
dimension zero. The following  \emph{twisted} stress-energy tensor~\cite{B&O} 
has that property:
\beq
T^{\textrm \tiny{impr}} =   T^{sug} - \de J^3 ,\qquad \qquad T^{\textrm
\tiny{impr}}(z) J^-(w) \sim \frac{\de J^-(w)}{z-w}.
\eeq
This assigns dimension zero, one and two to $J^-$, $J^3$, $J^+$, 
respectively\footnote{Notice,
however, the appearance of a central term of the form $-\frac{3k}{(z-w)^3}$ in
the $T^{impr}J^3$ OPE, which makes $J^3$  a \emph{quasi}-primary field.}.
In terms of the elementary  fields we have:
\bea\label{impr}
T^{\textrm \tiny{impr}} &=&  -\de \b \g +  \frac{1}{2} (\de \r)^2 - \a_\r \de^2
\r + T_\c  + T_m \nn \\
&=& T_m   + 
\left[-\de \b \g + \frac{1}{2} (\de \r)^2  -  \frac{1}{2} (\de \c)^2
 -  \a_\r \de^2 \r +  i \a_\c \de^2 \c\right].
\eea
We see that w.r.t. $T^{\textrm \tiny{impr}}$, $\b$ has dimension zero, and $\g$
has dimension one. The background charge of  $\r$  and the central charge become
\beq\label{cimpr}
\a_\r =(k+1)/\sqrt{2k+4}, \qquad \qquad  c_{impr} = 2 + (1-12\a_\r^2).
\eeq
(Recall that, in our construction, $T_\c  + T_m$ contributes zero to the 
central charge). If we impose the constraints
\beq 
\b(z)=k  , \qquad \qquad \de\r=  \de\chi, \label{constraints}
\eeq
and we fix $k$ such that
\beq 
\a_\r = i\a_\c,\label{alfa}
\eeq
the term in brackets in the second line of Eq.~(\ref{impr})
disappears, and the improved stress
tensor reduces to the physical one, with central charge $c_m$. At this point,
the level $k$ of the $sl(2,R)$ algebra is no longer arbitrary: from 
Eqs.~(\ref{chi},\ref{alfa}) it follows that $k$ is fixed in terms of $c_m$ by
\beq \label{c(k)}
\frac{(k+1)^2}{k+2} = - \frac{c_m + 1}{6}.
\eeq
Notice that, if we start from a physical CFT with positive central charge, this
relation requires $k+2$ to be negative. This is rather natural: recall for
example that, in the case of pure gravity in $AdS_3$, $k= -l/4G_N$
\cite{banados}. In the semi-classical limit, in which $k$ is large, Eq.~(\ref{c(k)})
 gives $c_m \simeq -6k = 3l/2G_N$, which agrees with the formula
for the ``classical'' central charge of $AdS_3$ gravity found long ago by Brown
and Henneaux \cite{B&H}.

{}From the above discussion, it seems reasonable  that the constraints Eqs.
(\ref{constraints}) should be enough to project out all the additional
auxiliary fields we have introduced to build the $sl(2,R)$ currents. At this
level, to require Eq.~(\ref{alfa}) sounds rather arbitrary: after all, by
imposing a constraint we eliminate some degrees of freedom,
independently of the numerical parameters of the theory.  In the next Section,
we will see that Eq.~(\ref{alfa}) is essential if we want to
impose the constraints consistently, using the BRST method. 

\section{The BRST Charge}\label{cohomology}

In this Section we show how the  constraints can be imposed using the BRST
formalism, as  it was done in  \cite{B&O} in the free field case. 
It is useful to introduce the quantity $\a_0 =
-i\a_\r$, which is real when $k+2$ is negative, and change variables to
\bea\label{opex+x-}
 X^+(z)  = \frac{1}{\sqrt{2}}\left[\r(z) + \c(z)\right], \;\;&\;&\;\;
X^-(z)=\frac{1}{\sqrt{2}}\left[ \r(z) - \c(z)\right], \nn\\
X^+(z)X^+(w) \sim 0, \qquad  X^+(z)X^-(w) &\sim& \ln (z-w), \qquad
X^-(z)X^-(w) \sim 0.
\eea
In terms of these variables, the second constraint in Eq.~(\ref{constraints})
reads $\de X^-(z) =0$, and the part of $T^{impr}$ depending on 
scalar fields is
\beq\label{tx+x-}
T^{impr}[X^+, X^-] = \de X^+ \de X^- -i \frac{\a_0 + \a_\c}{\sqrt{2}} \de^2 X^-
- i \frac{\a_0 -\a_\c}{\sqrt{2}} \de^2 X^+.
\eeq
Since we have two bosonic constraints we introduce two independent sets of
fermionic ghosts $(b,c)$ and $(B,C)$, of
conformal weights $(0,1)$, and $(1,0)$, respectively. Their OPEs are
\bea\label{opeghosts}
b(z)c(w) \sim \frac{1}{z-w}, &\qquad& c(z)b(w) \sim \frac{1}{z-w}, \nn\\
B(z)C(w) \sim \frac{1}{z-w}, &\qquad& C(z)B(w) \sim \frac{1}{z-w}.
\eea
Then, we define the total stress-energy tensor
\beq \label{total}
T^{tot}(z)=T^{impr}(z) + \de b(z) c(z) + \de C(z) B(z).
\eeq
Each set of  ghosts contribute $(-2)$ to the central charge,  so 
from Eq.~(\ref{cimpr}) the theory now has
\beq \label{ctot}
c_{tot}=c_{impr}- 4  = -1 + 12\a_0^2.
\eeq
Next, we define the following BRST current and charge:
\beq \label{BRS}
j_{B}(z) = [\b(z) -k]c(z) + \de X^-(z)C(z) , \qquad Q_{B}=\oint \frac{d
z}{2\pi i}j_{B}(z).
\eeq
The first term is the same one that was used in~\cite{B&O} to impose the
constraint $J^-=1$.
The charge $Q_{B}$  is nilpotent, since  the two terms anticommute with each
other, and have regular OPE with themselves. However,  $Q_{B}$ is \emph{not} 
conserved, in general: as one can see using 
Eqs.~(\ref{opex+x-},\ref{tx+x-},\ref{opeghosts}), the OPE between the stress 
energy tensor and the BRST current is
\beq\label{anomaly}
 T^{tot}(z)j_{B}(w) \sim  \frac{j_B(w)}{(z-w)^2} + \frac{\de j_B(w)}{z-w} + i
\sqrt{2}\left(\a_0 - \a_\c\right) \frac{C(w)}{(z-w)^3}.
\eeq
So, $Q_B$ does not commute with $T^{tot}$; we find instead 
\beq
\left[T^{tot}(z), Q_B\right] = -\frac{i}{\sqrt{2}}\left(\a_0 -
\a_\c\right)\de^2 C(z).
\eeq
Therefore, requiring that $Q_B$ is conserved forces us to impose  $\a_0
=\a_\c$\footnote{One may ask what happens, if we try to impose a more general
linear constraint, involving $X^+$ and $X^-$, with a term in $j_B$ of the form
$(a \de X^+ + b \de X^-)C$. It is easy to see that the nilpotency of $Q_B$
requires that 
either $a$ or  $b$ vanishes, and then its conservation implies that 
$\a_0 = \pm \a_\c$. The two choices yield the same physical spectrum.}. 

With this definition of the BRST charge, the total stress tensor of the theory
is physically equivalent to that of the original CFT, $T_m$. As a first
consequence of Eqs. (\ref{ctot},\ref{chi}), and since $\a_0 =\a_\c$, 
we have
$c_{tot}=c_m$. Moreover,  $T^{tot}=T_m$ modulo a BRST-exact operator:
\bea \label{exact}
T^{tot}-T_m & = &    \de X^+ \de X^- -i \frac{2 \a_0}{\sqrt{2}} \de^2 X^- -\de
\b \,\g  + \de b \,c + \de C \,B \nn \\
&=& - \left\{Q_{B}, \,\g \,\de b + i \sqrt{2} \a_0 \de B -(\de X^+)
B\right\}.
\eea
This shows that the physical CFT, and the extended theory with $sl(2,R)$
symmetry after BRST-projection, share the same stress tensor. 
In the next Section we will show that they also 
have the same spectrum of physical states.

\section{BRST Cohomology}

The BRST operator defined in the previous Section defines a cohomology on the
Hilbert space of the full $sl(2,R)$ theory. In this Section we show that this
cohomology is isomorphic to the space of states of the physical CFT. 

The Hilbert space of the $sl(2,R)$ theory (henceforth referred to as ${\cal
H}_{SL_2}$) is the tensor product of the original CFT Hilbert space ${\cal
H}_m$ with the free-field Fock spaces of the fields $\c$ and $\rho$, and of 
the pairs $(\b,\g)$, $(b,c)$, and $(B,C)$:
\beq\label{product}
 {\cal H}_{SL_2} =  {\cal H}_m  \,\otimes \,  {\cal H}_{\b,\g} \,\otimes \,
{\cal H}_{b,c}\, \otimes \,  {\cal H}_\r \, \otimes \,  {\cal H}_\c \, \otimes
\,  {\cal H}_{B,C}.
\eeq
The space of physical states ${\cal H}_{phys}$ is defined as the
$Q_B$-cohomology on ${\cal H}_{SL_2}$, referred to as  $H(Q_B)$. This is the
space of states annihilated by $Q_B$, modulo exact states, i.e. states 
belonging to the image of $Q_B$. 
The BRST operator is the sum of two terms, each of which acts  on different
factors in the tensor product in Eq.~(\ref{product})
\bea\label{Q_1+Q_2}
Q_B &=& \hat{Q}_1+\hat{Q}_2=Q_1\otimes 1_2  + (-)^{F_1}\otimes Q_2, \nn\\ 
  Q_1&=& \oint\frac{d z}{2\pi i} [\b(z) -k]c(z), \qquad 
Q_2 = \oint\frac{d z}{2\pi i}\de X^-(z)C(z).
\eea
$Q_1$ acts nontrivially only on 
${\cal H}_1\equiv  {\cal H}_{\b,\g} \,\otimes
\,  {\cal H}_{b,c}$, and  as the identity on the rest. $Q_2$ acts only on
${\cal H}_2 \equiv   {\cal H}_\r \otimes   {\cal H}_\c  \otimes {\cal
H}_{B,C}$. $(-)^{F_1}$ denotes the fermion parity  on ${\cal H}_1$.
Clearly $\hat{Q}_1$ and $\hat{Q}_2$ anticommute, and are separately nilpotent.
{}From a general cohomology-theoretical result, 
it follows that, in this
situation, the total cohomology is the tensor product of the two:
\beq
 H(\hat{Q}_1+\hat{Q}_2,{\cal H}_1\otimes{\cal H}_2) =  H(Q_1,{\cal H}_1) \otimes  
H(Q_2,{\cal H}_2).
\eeq
This statement is known in algebraic geometry as K\"unneth's formula  
(see e.g. \cite{griffiths}). For the benefit of the reader, we present a
proof adapted to our case in Appendix B. Clearly, since both $Q_1$ and 
$Q_2$
act as the identity operator on ${\cal H}_m$, the latter will be part of the
cohomology as a separate factor. Therefore the space of physical states has the
form
\beq
{\cal H}_{phys} = {\cal H}_m \otimes H(Q_1,{\cal H}_1) \otimes  
H(Q_2,{\cal H}_2).
\eeq
In what follows we show that both $H(Q_1)\equiv H(Q_1,{\cal H}_1)$ and 
$H(Q_2)\equiv H(Q_2,{\cal H}_2)$ are essentially
one-dimensional, so that the
spectrum of the $sl(2,R)$ theory after BRST-projection reduces to the spectrum
of the original CFT.

\subsection{$Q_1$ Cohomology}

Consider  the Hilbert space ${\cal H}_1$. This is the Fock space of oscillators
$(\b_n,\g_n, b_n, c_n)$, defined by the expansions
\bea\label{modes1}
\b(z) = \sum_{n} \frac{\b_n}{ z^{n}}, \qquad &&\qquad  \g(z) = \sum_{n} \frac{
\g_n}{z^{n+1}}, \nn\\
b(z) = \sum_{n} \frac{ b_n}{ z^{n}}, \qquad &&\qquad  c(z) = \sum_{n}
\frac{c_n}{z^{n+1}},
\eea
with non-vanishing (anti)commutation relations
\beq\label{comm1}
\left[\b_n, \g_m\right] = \delta_{n+m,0}, \qquad  \left\{c_n, b_m\right\} =
\delta_{n+m,0}.
\eeq
The space ${\cal H}_1$ contains  a vacuum state $|0\rangle\equiv |0\>_{\b,g}
\otimes |0\>_{b,c}$
satisfying
\bea
&&\b_n |0\> = b_n |0\> = 0, \qquad \qquad n\geq0, \nn\\
&&\g_n |0\> = c_n |0\> = 0, \qquad \qquad n\geq1.
\eea
All other states are built by repeatedly applying the remaining operators on
the vacuum. \\
In terms of oscillators, the BRST operator $Q_1$ reads
\beq\label{Q1}
Q_1=\sum_{n} \left(\b_n - k\,\delta_{n,0}\right) c_{-n}.
\eeq
Physical states are $Q_1$-closed,  ($Q_1 |\Phi\rangle = 0$), with two states
$|\Phi\>$ and $|\Phi'\>$ being equivalent if their difference is $Q_1$-exact,
($|\Phi\> = |\Phi'\> + Q_1 |\Psi\>$ for some $|\Psi\>$).
The vacuum state is not closed, since $Q_1 |0\> = (\b_0 -k)c_0 |0\> = -k\, c_0
|0\>$. To get a closed state  we must apply some $\g_0$ oscillators. This does 
not change the energy, since the Virasoro operator $L_0$ does not contain the
zero modes $\b_0$ and $\g_0$ [see Eq.~(\ref{impr})]:
\beq
L_0 = \sum_{n} n :\b_{-n}\g_{n}:+...,
\eeq
In particular,  we can take  as the physical
vacuum in the $(\b\g)$ sector the closed state $e^{k\,\g_0} |0\>$.  
It turns out that this vacuum state  is the only nontrivial state in the $Q_1$
cohomology. This is because~\cite{B&O} the pairs $(\b-k,\g)$ and
$(b,c)$ constitute a Kugo-Ojima quartet \footnote{A  Kugo-Ojima quartet is a
set of four fields that realize the following representation of the BRST
algebra:
\begin{displaymath}
[Q, b\} = \b, \qquad  [Q, \b\} = 0, \qquad  [Q, \g\} = c, \qquad [Q, c\} = 0.
\end{displaymath}}
\cite{K&O},  which is always projected out of the physical Hilbert
space $H(Q_1)$. To see this in our case, one recursively constructs 
the following set of projection operators:
\bea \label{projectors}
P^{(0)} &=& \exp (k\,\g_0) |0\>\<0|, \nn\\
P^{(N)} &=& {1\over N} \sum_{n\geq1} \left( b_{-n} P^{(N-1)} c_{n} -  \b_{-n}
P^{(N-1)} \g_{n}\right) \nn\\
&+& {1\over N} \sum_{n\geq0} \left[ c_{-n} P^{(N-1)} b_{n} + \g_{-n} P^{(N-1)}
\left(\b_{n}-k\,\delta_{n,0}\right)\right].
\eea
The operators  $P^{(N)}$ project on subspaces containing $N$ excitations of the
modes of $\b, \g, b,c$.  They commute with $Q_1$, and constitute a complete set
on $Ker\, Q_1$:  $\sum_N P^{(N)} = {\mathbf 1}_{Ker\, Q_1}$. Moreover, for $N\geq
1$ they are $Q_1$-exact:
\bea
P^{(N)} &=& \left\{Q_1, R^{(N)}\right\}, \nn\\
R^{(N)} &=& -{1\over N} \sum_{n\geq1}  b_{-n} P^{(N-1)} \g_{n} + {1\over
N}\sum_{n\geq 0} \g_{-n} P^{(N-1)} b_{n}.
\eea
Therefore, any closed state $|\Psi\>$ can be written as
\bea
|\Psi\> &=& \sum_{N\geq0} P^{(N)} |\Psi\> = P^{(0)} |\Psi\> + \sum_{N\geq
1}\left\{Q_1, R^{(N)}\right\}  |\Psi\>  \nn\\
 &=& \exp (k\,\g_0) |0\>\<0|\Psi\> + Q_1 \left(
\sum_{N\geq1}R^{(N)}|\Psi\>\right) .
\eea
So, the $Q_1$ cohomology reduces to the one-dimensional subspace generated
by $\exp (k\g_0) |0\>$, i.e. it  contains only the vacuum state:
\beq
H(Q_1) = \left\{\exp (k\,\g_0) |0\>_{\b,\g} \otimes |0\>_{b,c}\right\}.
\eeq

\subsection{$Q_2$ Cohomology} \label{Q-2cohom}

A similar discussion applies to $Q_2$. 
Consider now the Hilbert space ${\cal H}_2$. This is the Fock space of the 
oscillators 
$$(x^+, x^-, a^+_n, a^-_n, B_n, C_n),$$
defined by the expansions
\bea
X^+(z) &=& x^+ + a^+_0 \ln z - \sum_{n\neq 0} \frac{ a^+_n}{ n}z^{-n}, \qquad
X^-(z) = x^- + a^-_0 \ln z - \sum_{n\neq 0} \frac{ a^-_n}{n}z^{-n},  \label{modes2}\\
B(z) &=& \sum_{n} \frac{ B_n}{ z^{n+1}}, \qquad \qquad  C(z) \:=\: \sum_{n}
\frac{ C_n}{z^{n}},\label{roba1}  
\eea
with non-vanishing (anti)commutation relations
\bea
\left[a^+_n, a^-_m\right] &=& n \delta_{n+m,0}, \nn\\
\left[a^+_n, x^-\right] &=& \delta_{n,0}, \qquad \qquad \left[a^-_n,
x^+\right] \:=\: \delta_{n,0},  \nn\\
\left\{C_n, B_m\right\} &=& \delta_{n+m,0}. \label{comm2} 
\eea
The space ${\cal H}_2$ contains  a vacuum state $|0\rangle\equiv |0\>_{+,-}
\otimes |0\>_{B,C}$ annihilated by $a^{\pm}_n$, $n\geq 0$, by $C_n, B_n$, $n>0$
and by $B_0$. All other states are built by repeatedly applying the remaining
operators on the vacuum. 

In terms of oscillators, the BRST operator $Q_2$ reads
\beq\label{Q2}
Q_2=\sum_{n} a^-_n C_{-n}.
\eeq
The $Q_2$-cohomology is found with  the same  argument we used in the previous
Subsection for the $Q_1$-cohomology. The result is essentially the same, up to
the presence of the zero mode of the field $x^-$.
Indeed, the pairs of fields $(\de X^- , \de X^+)$ and $(B, C)$ constitute
another Kugo-Ojima quartet and, therefore, states created by their modes are
projected out of the cohomology. This time the  vacuum state is closed, so the
projector $P^{(0)}$ is just the projector on the vacuum of ${\cal H}_2$.
This leaves only the zero modes of $x^+$ and $x^-$ as possible candidates to
produce other physical states, besides the vacuum. Clearly any state of the
form $f(x^+,x^-) |0\>$ is  unphysical unless $f$ is independent of $x^+$, since
$[Q_2, f(x^+, x^-)] = C_0 [\de f(x^+,x^-)/\de x^+]$. However, there is no constraint
on the $x^-$ dependence. We can work with states with definite
$a^+_0$ and $a^-_0$ eigenvalues,  of the form $|p^+,p^-\> = \exp (p^+ x^- + p^- x^+)|0\>$. 
Physical states are required to have $p^-=0$. 
All  the states $|p^+,0\>$, with  arbitrary  $p^+$,  are closed, but 
they are not exact, and they are not even
BRST-equivalent to the zero-charge vacuum, since they are orthogonal to it.
However, they are all degenerate in energy with the vacuum. Indeed, 
from Eq.~(\ref{tx+x-}) we have
\beq
L_0^{tot} |p^+,0\> = \Big(a^+_0 a^-_0 + \, \textrm{{\small terms commuting
with} $x^-$} \Big) \exp (p^+ x^-) |0\> = p^+ a^-_0 |p^+,0\> =0.
\eeq
So, we can arbitrarily  choose any one of these states as ``the vacuum,'' and
all matrix element will be independent of this choice.

From the result of the last two Subsections, it follows that 
\beq
H(Q_B) \simeq {\cal H}_m \otimes \exp (k \g_0) |0\>_{aux}, 
\eeq 
where $|0\>_{aux}$ is the vacuum of 
the auxiliary fields we introduced in Section 3. 
This implies that  the physical Hilbert space $H(Q_B)$ can be identified 
with the Hilbert space ${\cal H}_m$ of the original CFT.
 
\section{Irreducible Representations}

In this Section we show how one can explicitly construct irreducible
representations of the $sl(2,R)$  affine Lie algebra generated by the currents
in Eq. (\ref{currents}), starting from a primary field of the ``matter'' CFT.
We focus on lowest weight representations of the current algebra, although
similar results hold for other types of representations.

A  lowest weight, irreducible representation is   realized in terms of a set of
\emph{affine primary} fields $H_{j,m}(z)$, ($m=j,j+1\ldots$) that obey
\bea
J^+(z)H_{j,m}(w) &\sim& - (m+j) \frac{H_{j,m+1}(w)}{z-w} \label{j+}, \\
J^3(z)H_{j,m}(w) &\sim&  \; m \;  \frac{H_{j,m}(w)}{z-w}   \label{j3}, \\
J^-(z)H_{j,m}(w) &\sim&  (m-j) \frac{H_{j,m-1}(w)}{z-w} \label{j-}.
\eea
Here the operator $H_{j,j}$ is the lowest weight operator. Each of these
operators is associated to a primary state, defined in the usual way as
$|j,m\>=H_{j,m}(0)|0\>$. If we expand the currents in modes,
\beq
J^a(z) = \sum_{n} \frac{J^a_n}{z^{n+1}},
\eeq
these states are annihilated by all positive modes, and form an irreducible
representation of the global $sl(2,R)$ subalgebra generated by the zero-modes
of the currents. The other states of the representation of the affine algebra
(the affine descendants) are obtained applying negative modes of the currents.

It turns out that
for  each \emph{Virasoro} primary ${\cal O}_h$ of the matter theory 
with stress energy tensor $T_m$, one can construct  one and  only one such 
irreducible representation.

Let us first describe how representations look like in the free
field  case, in which
the term  $T_m + T_\c$  in $J^+$ is absent (see e.g. ref.~\cite{gerasimov}). Clearly,  a lowest  weight operator
of charge $j$ under $J^3$ is  given by
\beq\label{jj}
H_{j,j}(z) =  \exp \left[\frac{2}{\a_+}\,j \r(z)\right].
\eeq
This satisfies Eqs. (\ref{j3}) and (\ref{j-}) with $m=j$. Now, commute 
repeatedly this operator with $J^+$, and use the OPE  to 
read off the other operators of the representation, 
from the right hand side of Eq.~(\ref{j+}).
The resulting primary fields are
\beq\label{jm}
H_{j,m}(z) = [\gamma(z)]^{m-j} \, \exp\left[ \frac{2}{\a_+} j
\r(z)\right], \qquad \qquad m=j,j+1,\ldots
\eeq

This procedure can be carried out in the general case in which the currents are
given by Eq.~(\ref{currents}). To see how, let us 
start with a primary operator of weight $\hat{h}$ under 
$T= T_m + T_\c$: ${\cal O}_{\hat{h}}(z)$. 
Define
\beq
H^{(\hat{h})}_{j,j}(z) = V^{(\r)}_j (z) {\cal O}_{\hat{h}}(z), \qquad  V^{(\r)}_j (z) = \exp
\left[\frac{2}{\a_+}\,j \r(z)\right].
\eeq
Clearly, this operator has vanishing OPE with $J^-$, and has charge $j$ 
under $J^3$, so it is a good candidate
for a lowest weight operator. On the other hand, its OPE with  $J^+$ reads:
\bea \label{j+1}
J^+(z) H^{(\hat{h})}_{j,j}(w) &\sim&  
-2 j \frac{\gamma(w)V^{(\r)}_j(w){\cal O}_{\hat{h}}(w)}{z-w} \nn \\
& +&  (k+2)\b^{-1}(z)V^{(\r)}_j(w)\left[\,\hat{h}\, \frac{{\cal O}_{\hat{h}}(w)}{(z-w)^2} 
+ \frac{\de{\cal O}_{\hat{h}}(w)}{z-w}\right].
\eea
This OPE has a second order pole \emph{unless} $\hat{h}=0$. So, in order for
$H^{(\hat{h})}_{j,j}$ to be an
affine primary field, the operator ${\cal O}_{\hat{h}}$ must have zero weight. 
This can be
achieved if we recall  that ${\cal O}_{\hat{h}}$ is a primary not only under $T_m$, 
 but also under $T_m+T_\c$. Now, given any primary operator 
of the \emph{matter} CFT,  ${\cal O}_h$,  of weight $h$ under $T_m$, we can always 
build another one, of  weight $\hat{h}=0$ under  $T_m+T_\c$, 
by dressing ${\cal O}_h$ with an
appropriate vertex operator involving $\c(z)$:
\beq
{\cal O}_0 = V^{(\c)}_q(z) {\cal O}_h, \qquad   V^{(\c)}_q(z) \equiv \exp [iq\c(z)],
\eeq
with $q$ satisfying
\beq\label{q(h)}
q(\frac{q}{2}-\a_\c)+h=0.
\eeq
Therefore, given any primary operator of arbitrary weight $h$ in the
matter theory, we can build a primary lowest weight operator for the
affine-Lie algebra,  
\beq\label{madd1}
H_{j,j}(w) = V^{(\r)}_j (z)V^{(\c)}_q(z) {\cal O}_h(z), 
\eeq
with $q$ and $h$
satisfying the relation Eq.~(\ref{q(h)}), and $j$ arbitrary.
 The next member of the representation, with $m=j+1$,  can be read off from the
right hand side of the OPE $J^+(z) H_{j,j}(w)$, and so on for all
$m=j+2,j+3,\ldots$. For this procedure to work, one has to check at every 
step that one does indeed obtain an affine  \emph{primary} operator on
the r.h.s. of the OPE with $J^+$, i.e. that every new operator one generates
has only first order poles in its OPE with the currents.  We argue that this is
the case as follows: Assume that the lowest weight
$H_{j,j}(z)$ is a primary. We can associate to it a primary \emph{state} in the
usual manner, by defining
$|j,j\> \equiv H_{j,j}(0) |0\>$. Since this state is primary, it is 
annihilated by all modes with $n$ strictly positive, 
in the mode expansion of the currents: $J^a(z) = \sum J^a_n z^{-n-1}$. Now, 
define the state
$|j,j+1\> = J^+_0 |j,j\>$. This is also  a primary  state, as one can check
using the affine-Lie algebra commutation relations written in terms of modes. 
We can associate to this primary state an  operator $H'(z)$, which is also a
primary, and by construction, is precisely the operator  $H_{j,j+1}(w)$
appearing on the r.h.s. of the OPE   $J^+(z) H_{j,j}(w)$. By applying  this
argument to  $H_{j,j+1}$, which now we know is a primary, we conclude
that  $H_{j,j+2}$ is a primary, and so on. Therefore, the only nontrivial
requirement is, that the lowest weight $H_{j,j}(z)$ is an  affine  primary.

As a check, we show explicitly that the second  operator in the ladder,
$H_{j,j+1}$ is indeed a primary, under the only assumptions that
$ H_{j,j}$ is a primary. We  can read off what the operator obtained by 
acting on $H_{j,j}$ with $J^+(z)$ is, from the residue of first order pole
in Eq. (\ref{j+1}):
\beq
H_{j,j+1}(z) = - \gamma V^{(\r)}_j V^{(\c)}_q {\cal O}_h  + \frac{k+2}{2j} \b^{-1}\, V^{(\r)}_j \,\de \left(V^{(\c)}_q {\cal O}_h\right).
\eeq
This clearly satisfies both Eqs. (\ref{j3}) and (\ref{j-}), with $m=j+1$. The
nontrivial part of the proof here, 
is to check that the OPE with $J^+$ has no poles of order
greater than one. A straightforward calculation shows that this is indeed the
case.

We have just seen that we can associate irreducible
representations of  ``angular momentum'' $j$ of the current algebra 
to each Virasoro primary state. Up to now,
there are no constraints on the value of $j$, since 
it does not  depend on $h$ in any
way. However, if we take into account the results of the previous Section,   we
see that,  for a generic value of $j$, an irreducible representation of the 
form described above will not contain any physical states. Indeed, 
because of the $Q_2$-cohomology conditions, 
these are constrained to have equal charge under $\de \r$ and $\de
\chi$. More explicitly, it is apparent from the above construction, that a
generic primary  state in a representation of type $j$ has the form
\beq
|\Psi\> = H  \exp \left[\frac{2}{\a_+}\,j \r(0)\right] \exp \left[i q
\c(0)\right] |0\>,
\eeq
where the operator $H$ does not contain exponentials of $\r$ and $\c$. 
Section \ref{Q-2cohom} taught us that, in order to be physical, the above state
must be proportional to $\exp (p^+x^-)|0\> = \exp [p^+(\r - \c)/\sqrt{2}]|0\>$, for
some $p^+$. This implies a relation between $j$ and $q$:\footnote{Recall that
in our notation $\a_+ = \sqrt{2k+4}$ is purely imaginary.}
\beq
j= \frac{\a_+ p^+}{2\sqrt{2}},   \qquad \qquad q = \frac{ip^+}{\sqrt{2}},
\eeq
or $q=2j/|\a_+|$. Using the relation between $q$ and $h$ given in 
Eq.~(\ref{q(h)}), we
see  that a representation with ``angular momentum'' $j$,
constructed over a matter primary of weight $h$, survives the BRST projection
iff
\beq \label{j(h)}
j(h)=\frac{|k+1|}{2} \left(1 - \sqrt{ 1 - 4 h \frac{|k+2|}{(k+1)^2}}\right).
\eeq
Here, we used the relation $\a_\c = \a_0 = |k+1|/|\a_+|$, and we took  the
smallest root of the quadratic equation for $j$.
We made this choice since it is natural to
associate the primary field with $h=0$ --the identity operator in the matter 
CFT-- with the trivial representation of $sl(2,R)$, which has $j=0$. 
No problem arises in
restricting  the cohomology to states which satisfy Eq.~(\ref{j(h)}). Indeed,
there are two conserved charges in our model, which commute with $Q_B$, namely,
$L_0$ and $a^+_0$. The latter is BRST-equivalent to a multiple of the 
$j$-charge. Therefore, we can always restrict the cohomology to a subspace of 
the full Hilbert space, on which these two charges obey some relation, e.g. 
Eq.~(\ref{j(h)}).
With this choice, every 
irreducible representation of the Virasoro algebra of the matter CFT, with
lowest weight $h$, can be promoted to the (unique) representation of the 
affine-Lie algebra, with lowest weight $j(h)$, given by Eq.~(\ref{j(h)}). 
The primary state $|h\> \in {\cal H}_m$ is identified with the physical
state $|\Psi(h)\> \in {\cal H}_{SL_2}$ given by:
\bea
|\Psi(h)\> & = & |h\>_m \otimes |p^+(h) , p^-=0
\>_{+,-} \otimes (\exp {k\,\g_0}) |0\>_{\b,\g} \otimes |0\>_{B,C,b,c}\,\,, 
\nn\\
p^+(h)&=& -i2(|k+2|)^{-1/2}j(h).
\eea
This is the converse of the result of~\cite{B&O}, in which it was shown that 
every irreducible  representation of the affine Lie algebra is  
BRST-projected onto a  single irreducible representation 
of the Virasoro algebra  in the physical Hilbert space. 
Notice that we do not claim that all affine descendants of $|h\>$ are physical;
quite the opposite: generically, only its {\em Virasoro} descendants are 
physical. One may wonder  if 
Eq.~(\ref{j(h)}) implies an upper bound on the possible values of $h$, 
which would  result in an ``exclusion principle'' similar to the
one  found in~\cite{sm}. This is true 
 if we require $j(h)$ to be real. However, this is not necessary: 
in  a lowest weight representation,  $j$ has to be real only if we 
restrict our attention  to \emph{unitary} representations of the 
global $sl(2,R)$ algebra, generated by the zero modes of the currents.
For a compact Lie group this, plus a condition on the weight (integrability), 
implies that the representation  of the full current algebra is unitary. 
Here, we need not impose such a restriction,
since we  only demand unitarity (i.e. positivity of the Hilbert space) in the
physical, BRST-reduced theory\footnote{Indeed, 
although the
global $sl(2,R)$ algebra admits (infinite-dimensional) unitary 
representations, the full $sl(2,R)$ current algebra does not~\cite{dixon}. 
Its representations contain at most a physical positive-definite subspace.}. 
This only requires that $h>0$, without any upper bound. 

Eq.~(\ref{j(h)}) may seem rather obscure, but its meaning becomes
more transparent in the semi-classical limit:  when $k$ is large,
 it reduces to $j \simeq h$. This is what we expect because 
we are dealing with a model that has two global $sl(2,R)$
structures, one generated by the zero modes of the affine currents, the other
by the Virasoro generators $L_1, L_0, L_{-1}$. While the latter has a clear
semi-classical meaning in terms of the gravitational 
interpretation of the model
--it generates the isometries of $AdS_3$-- the former does not.
Indeed, it must be
projected out of the physical space to give the theory a chance of possessing 
a metric formulation\footnote{The actions of CS theory and the Einstein-Hilbert
action coincide, yet, it is not true that \emph{every} gauge field 
configuration corresponds to a reasonable classical space-time: 
for this to be true the dreibein must
be invertible. Therefore, it is
reasonable to expect that, without additional boundary conditions,
as the ones described in Section (\ref{ads3 grav}), there may not be a
(semi-classical) gravitational interpretation of the model.}. Consider, 
however, the case in which some non-dynamical point-like sources are turned 
on inside $AdS_3$. Their energy fixes the value of the Casimir operator of
the Virasoro algebra of asymptotic isometries:  
$1/2(L_1 L_{-1} + L_{-1} L_{1} - 2 L_0^2) = -h(h-1)$. On the other
hand, we can interpret the same point source as a puncture in the disk, with an
associated holonomy of the gauge field. As  discussed in~\cite{seiberg}, 
this gives rise  to a representation of the $sl(2,R)$ current algebra based on
a lowest weight representation of the global $sl(2,R)$, with 
weight $h$. Therefore, 
at least for this kind of configurations, we do have $j=h$.

\section{Conclusions}

In this paper, we have found that all 2-d CFTs possess a hidden $sl(2,R)$ 
affine symmetry. This hidden symmetry is realized by embedding the CFT into a
new CFT, which contains more degrees of freedom and more states. It also 
contains a physical subspace, defined by a BRST cohomology, where it 
coincides with the original theory. 

One aim of our investigation was to extend the ``hyperholographic'' 
correspondence between 3-d pure gravity and 2-d Liouville theory to more 
general CFTs. This may allow for a non-perturbative definition of the 
Wheeler-De Witt wave function in any consistent 3-d quantum gravity coupled 
to matter.
Before achieving this goal, we should be able to find a wave function that 
obeys an acceptable Wheeler-De Witt equation. The problem is that 
Eq.~(\ref{m12}), and all its variations, 
do not look as yet physically acceptable. 
Specifically, the Gauss law one obtains,
\beq\label{mmm1}
F_{z\bar{z}}^a(z,\bar{z})\Psi=\rho^a(z,\bar{z})\Psi,
\eeq
has the correct dependence on the CS gravity fields, 
but not on the matter fields.
The charge density $\rho^a$, indeed, is essentially the stress-energy tensor 
of matter, so it must contain a piece quadratic in the the conjugate momenta 
of the matter fields. What we obtain by ``dressing'' CFT operators as in 
Section 7, instead, is linear in the matter fields and their 
conjugate momenta.

More modestly, though, our construction may be relevant to another problem 
of 3-d quantum gravity.

Many, in recent years, have conjectured that the microscopic origin of 3-d
black-hole entropy may be understandable in pure 3-d gravity.
Evidence for and against this idea can be found e.g. in~\cite{c}. 
The rationale for the
conjecture is that the CFT microstates that make up the black hole macrostate 
may be determined by internal properties of the Liouville theory itself.
What we know for certain is that these states cannot be the standard
normalizable states of the Liouville theory. A recent proposal to identify
these states has been put forward in~\cite{yc}. 

If this conjecture holds, and if we can identify the local operators that
correspond to black hole microstates,
then the quantum wave function of the black hole {\em is} given by 
Eqs.~(\ref{m13},\ref{mm5}). Now, the operators that appear
in Eqs.~(\ref{m13},\ref{mm5}) are BRST invariant combinations obtained by 
appropriately dressing the matter CFT operators. 
Furthermore, the matter CFT is standard: it is unitary,
and it has an $SL(2,C)$ invariant vacuum. So, unlike in the Liouville theory,
we expect to 
encounter no ambiguity in identifying the black hole microstates: they are
those of the matter CFT, dressed with our auxiliary ghost fields according 
to e.g. Eq.~(\ref{madd1}). So, we expect no ambiguity in
using Cardy's formula~\cite{car} to compute the asymptotic density of states. 

This and other questions raised by this paper are worth of future 
investigation. 

\subsection*{Acknowledgments}
We would like to thank L. Alvarez-Gaum\'e and 
P.A. Grassi for useful discussions.
M.P. is supported in part by NSF grants PHY-0245069 and PHY-0070787. F.N. 
is supported by an NYU Henry~M.~McCracken Fellowship. 

\section*{Appendix A: $AdS_3$ Boundary Conditions}
\setcounter{equation}{0}
\renewcommand{\theequation}{A.\arabic{equation}}
Here, we review  what the  $AdS_3$ asymptotic behavior of the metric 
means in terms of
the CS formulation, and what constraints it imposes on the boundary
affine $sl(2,R)$ currents.

The conditions defined in \cite{B&H} for a metric to be asymptotically  $AdS_3$
read:
\bea\label{metric}
{d s^2 \over l^2} &=&  \left[\left(1 + O(e^{-2 r})\right) d r^2  \,+
\,\left(e^{2 r}+O(1)\right) d x^+ d x^-\right. \nn\\
&+& \left.O(1)(d x^+)^2 \,+\, O(1)(d x^-)^2 + O(e^{-2 r})d rd x^+  +
O(e^{-2 r}) d rd x^-\right].
\eea
We parametrize the terms up to $O(e^{-r})$ by three $r$-independent functions
$F(x^+,x^-)$, $L(x^+,x^-)$, $\tilde{L}(x^+,x^-)$ in the following way:
\beq
 {d s^2 \over l^2} = d r^2 + \left(e^{2r} + F \right)\,d x^+ d x^- +
L\,(d x^+)^2 + \tilde{L}(d x^-)^2 +  O(e^{-2 r}).
\eeq
This implies that the dreibein 1-forms  $\{e^+, e^-, e^3\}$ are 
\bea\label{dreibein}
{1\over l}e^+  &=& e^r d x^+ \, + \, e^{-r} \left({1\over 2}Fd x^+ +
\tilde{L} d x^-\right), \\
{1 \over l} e^-&=& e^r d x^- \, + \, e^{-r} \left({1\over 2}Fd x^- + L d
x^+\right),\\
{1 \over l}e^3 &=& d r, 
\eea
up to terms of $O(e^{-2r})$. Since we have
\beq
d s^2  = (e^3)^2 + {1\over 2}\left(e^+ e^- + e^-e^+\right),
\eeq
we see that the flat metric used to raise and lower flat indexes is
\begin{displaymath}
\eta = \left( \begin{array}{ccc} 0 & 1/2 & 0 \\
				 1/2 & 0 & 0 \\
				 0 & 0 & 1  \\
\end{array} \right).
\end{displaymath}

Next, we want to find a set of spin connection 1-forms compatible with the
requirement that the torsion vanishes asymptotically. As the  three independent
components we choose $\om^{+3}, \om^{-3} , \om^{+-}$. We need  the vanishing
torsion equation, 
\beq \label{notorsion}
d e^a + \om^a_b \w e^b = 0,
\eeq
to be true at least up  to order $O(e^{-r})$. Requiring the $O(e^r)$ terms to
vanish fixes the leading order terms in $\om^{ab}$:
\beq
\om^{+3} = e^r d x^+ + O(e^{-r}), \quad  
\om^{-3} = e^r d x^- + O(e^{-r}), \quad 
\om^{+-} =  O(e^{-r}).
\eeq
We can parametrize the $O(e^{-r})$ terms by the functions $f(x^+,x^-)$,
$g(x^+,x^-)$, $h(x^+,x^-)$, $k(x^+,x^-)$, $p(x^+,x^-)$, $q(x^+,x^-)$ and the
1-form $\chi(x^+,x^-)$:
\bea
\om^{+3} &=&  e^r d x^+ + e^{-r} \left( h d x^+ + k d x^-  + p d r\right), \\
\om^{-3} &=&  e^r d x^- + e^{-r} \left( f d x^+ + g d x^-  + q d
r\right), \\
\om^{+-} &=&  e^{-r} \chi,
\eea
up to terms of $O(e^{-2r})$. Eqs. (\ref{notorsion}) then read
\bea
d e^+ + \om^+_3 \w e^3 +\om^{+}_+ \w e^+ &=&  {1\over2}\chi \w d x^+ +
O(e^{-r}), \\
d e^- + \om^-_3 \w e^3 +\om^{-}_- \w e^- &=& -{1\over2}\chi \w d x^- +
O(e^{-r}),\\
d e^3 + \om^3_+ \w e^+ + \om^3_-  \w e^- &=& -{1\over2} \left(g d x^- \w
d x^+ + q d r \w d x^+ \right) \nn\\
&&  -{1\over2} \left(h d x^+ \w d x^- + p d r \w d x^- \right) +
O(e^{-r}).
\eea
Requiring the $O(1)$ terms to vanish fixes $\chi=0$, $p=q=0$ and $g=h$.
Defining $\om_a = -1/2 \, \e_{abc} \om^{bc}$, we have\footnote{Remember  that
$\e_{+-3} = 1/2$, $\eta_{+-}=1/2$, $\eta^{+-}=2$, so that
$ \om^- = 2 \om_+ = 2\left(- 1/2  \om^{-3}\right)= -\om^{-3},$ $\om^+ = 2 \om_-
=2 \left( 1/2  \om^{+3}\right)= \om^{+3}.$}:
\bea\label{spinconn}
\om^+ &=& e^r d x^+ + e^{-r} \left( h d x^+ + k d x^-\right) +
O(e^{-2r}), \\
\om^- &=& - e^r d x^- - e^{-r} \left( h d x^- + f d x^+ \right) +
O(e^{-2r}),\\
\om^3 &=& 0 + O(e^{-2r}). 
\eea
The $sl(2,R)$ gauge  connection  1-forms are defined by the relations
\beq
A^a = {1\over l}\,e^a \, + \, \om^a, \qquad  
\tilde{A}^a = -{1\over l}\,e^a \, + \, \om^a,
\eeq
so from Eqs. (\ref{dreibein}) and (\ref{spinconn}) we get
\bea
A^+ &=& 2 e^r d x^+ + e^{-r} \left[\left({1\over 2} F + h\right) d x^+ +
\left(\tilde{L} + k\right)d x^- \right] +   O(e^{-2r}), \label{gaugeleft1} \\
A^- &=&  e^{-r} \left[\left({1\over 2} F - h\right) d x^- + \left(L -
f\right)d x^+ \right] +   O(e^{-2r}),\label{gaugeleft2}  \\
A^3 &=& d r  +   O(e^{-2r}),\label{gaugeleft3}
\eea
and
\bea
\tilde{A}^+ &=&  e^{-r} \left[\left(-{1\over 2} F + h\right) d x^+ +
\left(-\tilde{L} + k\right)d x^- \right] +   O(e^{-2r}), \label{gaugeright1}\\
\tilde{A}^- &=& -2 e^r d x^- -  e^{-r} \left[\left({1\over 2} F + h\right) d
x^- + \left(L + f\right)d x^+ \right] +   O(e^{-2r}), \label{gaugeright2}\\
\tilde{A}^3 &=&  -d r  +   O(e^{-2r}).\label{gaugeright3}
\eea
{}From the above equations we see that we 
need\footnote{From now on the  upper index always
refers to Lie algebra, the lower one to space-time coordinates.}  $A_-$ and
$\tilde{A}_+$  to vanish at the boundary as $O(e^{-r})$. This justifies the
choice  $A_-=\tilde{A}_+=0$ as boundary conditions on the gauge fields to make
the CS action differentiable.

Next, we  fix the gauge to set  $A_-=\tilde{A}_+=0$ everywhere in the 3-d
bulk, and not only on the boundary.
This means that we can choose $h=F/2$, $f=-L$, $k=-\tilde{L}$. Then, 
the connections become
\bea
A^+ &=& 2 e^r d x^+ + e^{-r} F  d x^+  +   O(e^{-2r}),
\label{gaugeleftfixed1} \\
A^- &=&  e^{-r} 2 L d x^+ \ +   O(e^{-2r}),\label{gaugeleftfixed2}  \\
A^3 &=& d r  +   O(e^{-2r}),\label{gaugeleftfixed3}
\eea
and
\bea
\tilde{A}^+ &=&  -e^{-r}2\tilde{L} d x^-  +   O(e^{-2r}),
\label{gaugerightfixed1}\\
\tilde{A}^- &=& -2 e^r d x^- -  e^{-r} F  d x^-  +   O(e^{-2r}),
\label{gaugerightfixed2}\\
\tilde{A}^3 &=&  -d r  +   O(e^{-2r}).\label{gaugerightfixed3}
\eea

Now we go to the WZW description to find out what consequences  Eqs.
(\ref{gaugeleftfixed1}) through (\ref{gaugerightfixed3}) have on the affine currents. The
 WZW variables are $SL(2,R)$ group elements $U(x^\mu)$, $\tilde{U}(x^\mu)$
defined by the relations
\bea
A_r = U \de_r U^{-1}, \qquad &&\qquad    A_\phi = U \de_\phi U^{-1},\\
\tilde{A}_r = \tilde{U} \de_r \tilde{U}^{-1}, \qquad &&\qquad    \tilde{A}_\phi =
\tilde{U} \de_\phi \tilde{U}^{-1}.
\eea

These variables are not well-suited to define the $sl(2,R)$ currents, since 
when Eqs. (\ref{gaugeleftfixed1}) through
(\ref{gaugerightfixed3}) are satisfied, the $\pm$ components of $A$ and
$\tilde{A}$ either vanish or diverge on the boundary. 
Therefore, we write  $U$ and $\tilde{U}$ as
\beq\label{g}
U(x^\mu) = \exp\{r t^3\}g (x^\mu), \qquad \tilde{U}(x^\mu) = 
\exp\{-rt^3\}\tilde{g} (x^\mu).
\eeq
and construct the  currents  with the group elements $g$ and $\tilde{g}$: 
\beq\label{well-def}
  J^a(x^+,x^-) = \lim_{r\to\infty}k \tr[ t^a g\de_\f g^{-1}], \qquad
\tilde{J}^a(x^+,x^-) = \lim_{r\to\infty} k \tr [ t^a \tilde{g}\de_\f
\tilde{g}^{-1}].
\eeq
They are related to the boundry values of the 
gauge fields by\footnote{We make use of the relations
$ \exp\{r t^3\} t^{+}\exp\{-r t^3\} = (\exp\{r\}) t^{+},$ $\exp\{r t^3\}
t^{-}\exp\{-r t^3\} = (\exp\{-r\}) t^{-}$.}
\bea
J^+(x^+,x^-) = \lim_{r\to\infty} {k\over 2}  e^r (A_+)^-, &&
\tilde{J}^+(x^+,x^-) =
\lim_{r\to\infty} {k\over 2}  e^{-r} (\tilde{A}_-)^-, \\
J^-(x^+,x^-) = \lim_{r\to\infty} {k\over 2}  e^{-r} (A_+)^+, &&
\tilde{J}^-(x^+,x^-) = \lim_{r\to\infty} {k\over 2}  e^r (\tilde{A}_-)^+,\\
J^3(x^+,x^-) = \lim_{r\to\infty} {k\over 2}  (A_+)^3,   
&&\tilde{J}^3(x^+,x^-) =
\lim_{r\to\infty} {k\over 2} (\tilde{A}_-)^3. 
\eea
{}From the above relations  and  Eqs. (\ref{gaugeleftfixed1}) through
(\ref{gaugerightfixed3}) we see that the currents in eq.~(\ref{well-def}) are well defined, and that they must satisfy
\beq
J^- = k, \qquad  J^3 = 0; \qquad\qquad  \tilde{J}^+ = -k,\qquad  \tilde{J}^3=0.
\eeq
On the other hand
\beq
J^+ = kL(x^+,x^-),  \qquad \tilde{J}^- =-k\tilde{L}(x^+,x^-),
\eeq
are arbitrary, and constitute the boundary degrees of freedom that survive
after enforcing the $AdS_3$ asymptotic conditions.

\section*{Appendix B: K\"unneth's Formula in BRST Cohomology}
\setcounter{equation}{0}
\renewcommand{\theequation}{B.\arabic{equation}}
Here we prove that the cohomology of $Q = Q_1 + Q_2$ is the direct product
of the cohomologies of $Q_1$ and $Q_2$.

Consider a Hilbert space ${\cal H} = {\cal H}_1 \otimes {\cal H}_2$, where
${\cal H}_1$ and ${\cal H}_2$ are Hilbert spaces with a ${\mathbf Z}_2$ grading
given by a fermion parity.
Consider  a nilpotent  operator $Q$ acting on ${\cal H}$, of the form  $Q = Q_1
\otimes 1_2 + (-)^{F_1} \otimes Q_2$, where $F_1$ is the Fermion
number
in the space ${\cal H}_1$, with $Q_1^2 = Q_2^2 = 0$. Then
we can consider the spaces $H(Q,{\cal H})= Ker\, Q / Im\, Q $, $H(Q_1,{\cal H}_1)=
Ker\, Q_1 / Im\,Q_1 $ and $H(Q_2,{\cal H}_2)= Ker\, Q_2 / Im\,Q_2 $, defined as the
spaces of equivalence classes $[\psi_\a]$, with  $\psi_\a \sim \psi_\a'$ if
$\psi_\a'=\psi_\a + Q_\a \xi$, where the index $\a$ refers to any of the three
pairs $({\cal H}, Q)$, $({\cal H}_1, Q_1)$,  $({\cal H}_2, Q_2)$.  If $Q_1$ and
$Q_2$ are hermitian operators, the cohomology  spaces carry a Hilbert space
structure inherited from that of ${\cal H}$, ${\cal H}_1$ and ${\cal H}_2$,
respectively.

We are going to prove the following \\

{\bf K\"unneth's Theorem:} The spaces $H(Q,{\cal H}_1\otimes {\cal H}_2)$ and
$H(Q_1,{\cal H}_1) \, \otimes \, H(Q_2,{\cal H}_2)$ are isomorphic as Hilbert
spaces.\\

We  prove this statement by constructing a map between the two spaces and then
showing that it is a unitary  isomorphism\footnote{We follow closely the proof
given in \cite{griffiths} for the analogous theorem  in the context of homology
of manifolds.}. The linear map in question is defined in a natural  way on
separable elements of $H(Q_1,{\cal H}_1) \, \otimes \, H(Q_2,{\cal H}_2)$ as
\bea
 &&\mu :  H(Q_1,{\cal H}_1) \, \otimes \, H(Q_2,{\cal H}_2) \longrightarrow
H(Q,{\cal H}_1\otimes {\cal H}_2)  \nn \\
 && \qquad \qquad \mu \Big([|\a\>] \otimes [|\beta\>]\Big) =
\Big[|\a\>\otimes|\beta\>\Big]. \label{mu}
\eea
The action of this map  depends only on the equivalence classes $[|\a\>]$, 
$[|\beta\>]$, not on the representatives $\a\in Ker\, Q_1$, $\b\in Ker\, Q_2$:
indeed, if $\a' \sim \a$, then
\bea
\Big[|\a'\>\otimes|\beta'\>\Big] &=& \Big[\left(|\a\> + Q_1 |\xi\>\right)
\otimes|\beta\> \Big] = \nn\\ &=& \Big[|\a\> \otimes|\beta\> +
Q\, \left( |\xi\> \otimes|\beta\>\right) \Big] =
\Big[|\a\>\otimes|\beta\>\Big].
\eea
Therefore $\mu$ is well defined as a function of the cohomology. To show that
$\mu$ is one-to-one and onto, 
we proceed as follows. Decompose the spaces ${\cal H}_1$
and  ${\cal H}_2$ as ${\cal H}_i = Ker\, Q_i \oplus N_i$, $i=1,2$,  where 
$N_i$ is
the orthogonal complement to  $Ker\, Q_i$. Introducing the projections $\pi_i$
defined as
\newpage
\bea
&&\pi_i : Ker\, Q_i \longrightarrow H(Q_i) \nn\\
&& \qquad  \pi_i (\psi_i) = [\psi_i],
\eea
so that $Ker\,\pi_i \equiv Im\,Q_i$   we can further write   
$Ker\, Q_i = Ker\, \pi_i \,\oplus \,V_i$
where $V_i$ is  naturally isomorphic to $Ker\, Q_i/Ker\, \pi_i =
H(Q_i)$, through the identification  $v_i \leftrightarrow [v_i]$. Therefore we
have
\bea \label{decomp0}
{\cal H}_i &=& Im\,Q_i \oplus V_i \oplus  N_i, \nn\\
   |\psi_i\> &=& |\sigma_i\> + |\tau_i\> + |\nu_i\>  \qquad \forall
\:|\psi_i\>\in {\cal H}_i,
\eea
with $|\sigma_i\>\in Im\,Q_i$, $|\tau_i\>\in V_i \subset Ker\, Q_i$, $|\nu_i\>\in
N_i$ uniquely specified by $\psi_i$.
Now, consider an element $|\Psi\>$ of $Ker\, Q \subset {\cal H}_1\otimes {\cal
H}_2$. According to the above decomposition, we have
\bea\nn
|\Psi\> &=&  \sum_k |\psi_1^k\> \otimes |\psi_2^k\> = \left(\sum_k
|\sigma_1^k\> \otimes |\sigma_2^k\> + \sum_k |\sigma_1^k\> \otimes  |\tau_2^k\>
+ \sum_k |\tau_1^k\> \otimes |\sigma_2^k\>\right) \nn\\
&+& \left( \sum_k |\tau_1^k\> \otimes |\tau_2^k\> + \sum_k |\tau_1^k\>\otimes
|\nu_2^k\> + \sum_k |\nu_1^k\>\otimes  |\tau_2^k\> + \sum_k  |\nu_1^k\>
\otimes|\nu_2^k\>\ \right) \nn \\
&+&\sum_k  |\nu_1^k\>\otimes|\sigma_2^k\> +  \sum_k |\sigma_1^k\> \otimes
|\nu_2^k\>.  \label{decomp}
\eea
The terms in the first line are $Q$-exact:
\bea
&&\sum_k |\sigma_1^k\> \otimes |\sigma_2^k\> + \sum_k |\sigma_1^k\> \otimes
|\tau_2^k\> + \sum_k |\tau_1^k\> \otimes |\sigma_2^k\> = \nn\\ &=& Q
\left(\sum_k |\tilde{\nu}_1^k\> \otimes  |\sigma_2^k\>
+\sum_k  |\tilde{\nu}_1^k\> \otimes  |\tau_2^k\> + \sum_k (-)^{F_1}|\tau_1^k\>
\otimes  |\tilde{\nu}_2^k\>\right),  \nn
\eea
where  $|\sigma_i^k\> = Q_i|\tilde{\nu}_i^k\>$. Moreover, the last term in Eq.
(\ref{decomp}) can be written as
\beq
\sum_k |\sigma_1^k\> \otimes |\nu_2^k\> =  Q \left( |\tilde{\nu}_1^k\>\otimes
|\nu_2^k\> \right) - \sum_k (-)^{F_1} |\tilde{\nu}_1^k\> \otimes
|\tilde{\sigma}_2^k\>,
\eeq
where $|\tilde{\sigma}_2^k\> = Q_2  |\nu_2^k\>$. Therefore $|\Psi\>$ is
$Q$-equivalent
to a $|\Psi'\>$, which is given by
\beq \label{decomp2}
|\Psi'\> =  \sum_k |\tau_1^k\> \otimes |\tau_2^k\> + \sum_k |\tau_1^k\>\otimes
|\nu_2^k\> + \sum_k |\nu_1^k\>\otimes  |\tau_2^k\> + \sum_k  |\nu_1^k\>
\otimes|\nu_2^k\> + \sum_{k'}|\nu_1^{k'}\> \otimes|\sigma_2^{k'}\> .
\eeq
When $|\Psi'\>$ is in $Ker\, Q$ we have
\bea
0 = Q |\Psi'\> &=&  \sum_k (-)^{F_1}|\tau_1^k\>\otimes |\tilde{\sigma}_2^k\> +
\sum_k |\tilde{\sigma}_1^k\>\otimes  |\tau_2^k\> \nn\\
&+& \sum_k  |\tilde{\sigma}_1^k\> \otimes|\nu_2^k\> + \sum_k
(-)^{F_1}|\nu_1^k\> \otimes  |\tilde{\sigma}_2^k\> +
\sum_{k'}|\tilde{\sigma}_1^{k'}\> \otimes|\sigma_2^{k'}\>.
\eea
Since all terms on the r.h.s. are linearly independent, this is possible
only if they vanish separately. This in turn implies that all terms in Eq.
(\ref{decomp2}) except the first one must vanish. For instance, it cannot
happen that  $\sum_k |\tau_1^k\>\otimes |\nu_2^k\> \neq 0$ but
$0=\sum_k |\tau_1^k\>\otimes |\tilde{\sigma}_2^k\> = Q_2 ( \sum_k
|\tau_1^k\>\otimes |\nu_2^k\>)$, since by construction  $\sum_k
|\tau_1^k\>\otimes |\nu_2^k\>$ is \emph{not} in $Ker\, Q_2$. Therefore, any
element in $|\Psi\> \in Ker\, Q$ is in the same cohomology class as an element of
the form
\beq \label{decomp3}
|\Psi'\> = \sum_k |\tau_1^k\> \otimes |\tau_2^k\>,  \qquad \qquad \textrm{for
some}\; \tau_i^k \in V_i \subset Ker\, Q_i.
\eeq
Moreover, $|\Psi\>$ is in $Im\,Q$ if and only if $|\Psi'\>=0$.
With this result, it is straightforward to show that the map $\mu$ defined in
Eq. (\ref{mu}) is one-to-one and onto. 
It is  obviously onto, since given $|\Psi\> \in
Ker\, Q$ we have
\beq
[|\Psi\>] = [|\Psi'\>] = \Big[\sum_k |\tau_1^k\> \otimes |\tau_2^k\> \Big] =
\mu \sum_k [|\tau_1^k\>] \otimes [|\tau_2^k\>],
\eeq
for some $\tau_i^k \in Ker\, Q_i$. To show that it is one-to-one,
take an element $\sum_k[|\a^k\>] \otimes [|\beta^k\>] \in   H(Q_1,{\cal H}_1)
\, \otimes \, H(Q_2,{\cal H}_2)$, which is mapped into the zero cohomology
class of  $H(Q,{\cal H})$. If  $[|\Psi\>] = \mu(\sum_k[|\a^k\>] \otimes
[|\beta^k\>])= \Big[\sum_k |\a^k\> \otimes |\beta^k\> \Big]= 0 \in  H(Q,{\cal
H})$, the representative in Eq. (\ref{decomp3}) vanishes. Moreover, 
thanks to the decomposition in 
Eq.~(\ref{decomp0}), for every $k$ in  $\sum_k |\a^k\> \otimes
|\beta^k\>$,  either  $|\a^k\>$ is  $Q_1$-exact or $|\beta^k\>$ is $Q_2$-exact,
so that $\sum_k[|\a^k\>] \otimes [|\beta^k\>] = 0 \in  H(Q_1,{\cal H}_1) \,
\otimes \, H(Q_2,{\cal H}_2) $.

Lastly, 
we show that the map $\mu$ is unitary. That is, given the scalar product
structures of $H(Q_1)$, $H(Q_2)$ and   $H(Q_1+Q_2)$\footnote{These are defined
by their value on representatives:
$\< [\psi], [\psi]' \> := \< \psi, \psi' \>$, which does not depend on the
choice of $\psi \in [\psi]$ and $\psi' \in [\psi]'$ since $Q$-exact states are
orthogonal to $Ker\, Q$.}, we have
\beq
\<\mu(\zeta), \mu(\zeta')\>_{H(Q)} =  \<\zeta, \zeta'\>_{H(Q_1)\otimes H(Q_2)}.
\eeq
This is straightforward: take $\zeta= \sum_k [\a_k] \otimes [\b_k]$, $\zeta'=
\sum_l [\a_l'] \otimes [\b_l']$, then from the definition of $\mu$
\bea
\<\mu(\zeta), \mu(\zeta')\> &=& \< \left[\sum_k \a_k \otimes \b_k\right],
\left[\sum_l \a_l' \otimes \b_l'\right]\> = \< \sum_k \a_k \otimes \b_k, \sum_l
\a_l' \otimes \b_l'\> \nn\\
&=&  \sum_{k,l} \<\a_k,\a'_l\> \<\b_k,\b_l'\>.
\eea
On the other hand:
\bea
\<\zeta, \zeta'\> &=& \< \sum_k [\a_k] \otimes [\b_k] , \sum_l [\a_l'] \otimes
[\b_l'] \> = \sum_{k,l}\<[\a_k],[\a'_l]\> \<[\b_k],[\b_l']\> \nn\\
&=&  \sum_{k,l} \<\a_k,\a'_l\> \<\b_k,\b_l'\>,
\eea
thus proving that $\mu$ is a unitary isomorphism.


\begin{thebibliography}{00}
\bibitem{'t} G.~'t Hooft,
arXiv:gr-qc/9310026. 
\bibitem{s} L.~Susskind,
J.\ Math.\ Phys.\  {\bf 36}, 6377 (1995)
[arXiv:hep-th/9409089].
\bibitem{b} J.~D.~Bekenstein,
Phys.\ Rev.\ D {\bf 7}, 2333 (1973).
\bibitem{agmoo} O.~Aharony, S.~S.~Gubser, J.~M.~Maldacena, H.~Ooguri and Y.~Oz,
Phys.\ Rept.\  {\bf 323}, 183 (2000)
[arXiv:hep-th/9905111].
\bibitem{m} J.~M.~Maldacena,
Adv.\ Theor.\ Math.\ Phys.\  {\bf 2}, 231 (1998)
[Int.\ J.\ Theor.\ Phys.\  {\bf 38}, 1113 (1999)]
[arXiv:hep-th/9711200].
\bibitem{sm} J.~M.~Maldacena and A.~Strominger,
JHEP {\bf 9812}, 005 (1998)
[arXiv:hep-th/9804085].
\bibitem{db} J.~de Boer,
Nucl.\ Phys.\ B {\bf 548}, 139 (1999)
[arXiv:hep-th/9806104].
\bibitem{B&H} J.~D.~Brown and M.~Henneaux,
Commun.\ Math.\ Phys.\  {\bf 104}, 207 (1986).
\bibitem{m2} J.~M.~Maldacena, talk at Strings 2002, Cambridge, UK;
G.~T.~Horowitz and J.~Maldacena,
arXiv:hep-th/0310281.
\bibitem{w1} E.~Witten,
Commun.\ Math.\ Phys.\  {\bf 121}, 351 (1989).
\bibitem{seiberg} S.~Elitzur, G.~W.~Moore, A.~Schwimmer and N.~Seiberg,
Nucl.\ Phys.\ B {\bf 326}, 108 (1989); G.~W.~Moore and N.~Seiberg,
Phys.\ Lett.\ B {\bf 220}, 422 (1989).
\bibitem{B&O} M.~Bershadsky and H.~Ooguri,
Commun.\ Math.\ Phys.\  {\bf 126}, 49 (1989).
\bibitem{superB&O}
M.~Bershadsky and H.~Ooguri,
Phys.\ Lett.\ B {\bf 229}, 374 (1989).

\bibitem{chvd} O.~Coussaert, M.~Henneaux and P.~van Driel,
Class.\ Quant.\ Grav.\  {\bf 12}, 2961 (1995)
[arXiv:gr-qc/9506019].
\bibitem{banados} M.~Banados,
arXiv:hep-th/9901148.
\bibitem{w} E.~Witten,
Nucl.\ Phys.\ B {\bf 311}, 46 (1988).
\bibitem{Achucarro:vz}
A.~Achucarro and P.~K.~Townsend,
Phys.\ Lett.\ B {\bf 180}, 89 (1986).
\bibitem{tv} V.~G.~Turaev and O.~Y.~Viro,
Topology {\bf 31}, 865 (1992).
\bibitem{sonnenschein} J.~Sonnenschein,
Nucl.\ Phys.\ B {\bf 309}, 752 (1988).
\bibitem{stone} M.~Stone,
Phys.\ Rev.\ Lett.\  {\bf 63}, 731 (1989).
\bibitem{HMS} M.~Henneaux, L.~Maoz and A.~Schwimmer,
Annals Phys.\  {\bf 282}, 31 (2000)
[arXiv:hep-th/9910013].
\bibitem{Forgacs:ac}
P.~Forgacs, A.~Wipf, J.~Balog, L.~Feher and L.~O'Raifeartaigh,
Phys.\ Lett.\ B {\bf 227}, 214 (1989).
\bibitem{hmtz} M.~Henneaux, C.~Martinez, R.~Troncoso and J.~Zanelli,
Phys.\ Rev.\ D {\bf 65}, 104007 (2002)
[arXiv:hep-th/0201170].
\bibitem{Wakimoto} M.~Wakimoto,
Commun.\ Math.\ Phys.\  {\bf 104}, 605 (1986).
\bibitem{zamol} A.~B.~Zamolodchikov, Montreal lectures, 1988, unpublished. 
\bibitem{Feigin}
B.~L.~Feigin and E.~V.~Frenkel, in \emph{Physics and mathematics of strings}, ed. L.~Brink  et al. (World Scientific, Singapore, 1990), 271-316;
Lett.\ Math.\ Phys.\  {\bf 19}, 307 (1990). 
\bibitem{gerasimov}
A.~Gerasimov, A.~Morozov, M.~Olshanetsky, A.~Marshakov and S.~L.~Shatashvili,
Int.\ J.\ Mod.\ Phys.\ A {\bf 5}, 2495 (1990).
\bibitem{Bernard:1989iy}
D.~Bernard and G.~Felder,
Commun.\ Math.\ Phys.\  {\bf 127}, 145 (1990).
\bibitem{bouwknegt}
P.~Bouwknegt, J.~G.~McCarthy and K.~Pilch,
Phys.\ Lett.\ B {\bf 234}, 297 (1990); 
Nucl.\ Phys.\ B {\bf 352}, 139 (1991);
P.~Bouwknegt, J.~G.~McCarthy, D.~Nemeschansky and K.~Pilch,
Phys.\ Lett.\ B {\bf 258}, 127 (1991).
\bibitem{ohta}
M.~Kuwahara, N.~Ohta and H.~Suzuki,
Nucl.\ Phys.\ B {\bf 340}, 448 (1990);  M.~Kuwahara, N.~Ohta and H.~Suzuki,
Phys.\ Lett.\ B {\bf 235}, 57 (1990); N.~Ohta and H.~Suzuki,
Nucl.\ Phys.\ B {\bf 332}, 146 (1990).
\bibitem{Furlan}
P.~Furlan, A.~C.~Ganchev, R.~Paunov and V.~B.~Petkova,
Nucl.\ Phys.\ B {\bf 394}, 665 (1993)
[arXiv:hep-th/9201080].
\bibitem{bouwknegtrev}
P.~Bouwknegt, J.~G.~McCarthy and K.~Pilch,
Prog.\ Theor.\ Phys.\ Suppl.\  {\bf 102}, 67 (1990).
\bibitem{Rasmussen} J.~L.~Petersen, J.~Rasmussen and M.~Yu,
Nucl.\ Phys.\ Proc.\ Suppl.\  {\bf 49}, 27 (1996)
[arXiv:hep-th/9512175].
\bibitem{griffiths} P.~Griffiths and 
J.~Harris, \emph{Principles of Algebraic Geometry}, Wiley \& Sons, 1978. 
\bibitem{K&O} T.~Kugo and I.~Ojima,
Prog.\ Theor.\ Phys.\  {\bf 60}, 1869 (1978).
\bibitem{dixon}
L.~J.~Dixon, M.~E.~Peskin and J.~Lykken,
Nucl.\ Phys.\ B {\bf 325}, 329 (1989).
\bibitem{c} S.~Carlip,
Rept.\ Prog.\ Phys.\  {\bf 64}, 885 (2001)
[arXiv:gr-qc/0108040]; 
Class.\ Quant.\ Grav.\  {\bf 15}, 3609 (1998)
[arXiv:hep-th/9806026].
\bibitem{yc} Y.~Chen,
arXiv:hep-th/0310234.
\bibitem{car} J.~L.~Cardy,
Nucl.\ Phys.\ B {\bf 270}, 186 (1986).

\end{thebibliography}
\end{document}